\documentclass[sort&compress,3p]{elsarticle}

\usepackage{lineno}
\usepackage{amsmath}
\usepackage{amsfonts}
\usepackage{amssymb}
\usepackage{mathrsfs}
\usepackage{graphicx}% Include figure files
\usepackage{subfigure}
\usepackage[dvipdfm]{hyperref}
\modulolinenumbers[5]

\begin{document}

\begin{frontmatter}

\title{Quantum evolution speed in the finite-temperature bosonic
	environment}
%\tnotetext[mytitlenote]{Fully documented templates are available in the elsarticle package on \href{http://www.ctan.org/tex-archive/macros/latex/contrib/elsarticle}{CTAN}.}

%% Group authors per affiliation:
\author{Jun-Qing Cheng}
\author{Guo-Qing Zhang}
\author{Jing-Bo Xu\corref{cor}}
\cortext[cor]{Corresponding author}
\ead{xujb@zju.edu.cn}
\address{Zhejiang Institute of Modern Physics and Physics Department, Zhejiang University, 310027 Hangzhou, People’s Republic of China}
%\fntext[myfootnote]{Since 1880.}

%% or include affiliations in footnotes:
%\author[mymainaddress,mysecondaryaddress]{Elsevier Inc}
%\ead[url]{www.elsevier.com}

%\address[mymainaddress]{1600 John F Kennedy Boulevard, Philadelphia}
%\address[mysecondaryaddress]{360 Park Avenue South, New York}

\begin{abstract}
We investigate  the  quantum evolution speed of a qubit in two kinds  of finite-temperature environments. The first environment is a bosonic bath with Ohmic-like spectrum. It is found that the high  temperature not only leads to the speed-up  but also  speed-down processes in the weak-coupling regime, which is different from the strong-coupling case where only exhibits speed-up process, and   the effects of  Ohmicity parameter of the bath on the quantum evolution speed are also different in the strong-coupling and weak-coupling regimes.  Furthermore, we realize the controllable and stationary quantum evolution speed by applying the bang-bang pulse.  For the second nonlinear bath, we study the quantum  evolution speed of a qubit by resorting to the hierarchical equations of motion method beyond the Born-Markov approximation.  It is shown that
the performances of quantum  evolution speed in  weak-coupling and strong-coupling regimes are also different. In particular, the  quantum evolution speed  can be decelerated by the rise of temperature in the strong-coupling regime which is an anomalous phenomenon and contrary to the common recognition that quantum evolution speed  always increases with the temperature.
\end{abstract}


\end{frontmatter}

%\linenumbers

\section{Introduction}
Quantum speed limit describes the maximal  evolution speed  of a quantum system from an initial state to a target state, which has been found  applications in the fields of quantum computation \cite{lloyd1999ultimate}, quantum thermodynamics \cite{PhysRevLett.105.170402}, quantum metrology \cite{demkowiczdobrzanski2012} and quantum control \cite{caneva2009optimal}. The minimum evolution time between two distinguishable states of a quantum system is defined as the quantum speed limit time (QSLT) \cite{PhysRevLett.65.1697,americanjournal1.16940,Margolus1998The,PhysRevA.67.052109,PhysRevLett.103.160502}, and has been  widely used to  characterize the maximal  evolution speed. The QSLT first proposed  for a closed quantum  system to naturally evolve to  an orthogonal state is characterized by unifying the   Mandelstam-Tamm (MT) bound \cite{J.Phys.USSR}  and Margolus-Levitin (ML) bound \cite{Margolus1998The}. Since the inevitable interaction of quantum systems with their surrounding environment, the generalizations of QSLT for open quantum systems have attracted much attention, and some valuable works have been done \cite{PhysRevLett.110.050402,PhysRevLett.110.050403,PhysRevLett.111,Zhang2014Quantum} in recent years. In Ref. \cite{PhysRevLett.111}, the authors have proposed  a unified bound of QSLT including both MT and ML types for non-Markovian dynamics with pure initial states.
For a wider range of applications, another unified bound of QSLT  applied to both mixed and pure initial states has been derived  by introducing relative purity \cite{Audenaert2014} as the distance measure \cite{Zhang2014Quantum}. These results have stimulated the interest of some further research about quantum speed limit.

Recently, some remarkable progresses  about  the analysis of  environmental effects on the   quantum speed limit for  open quantum systems have been made. For example,  in Refs. \cite{PhysRevLett.111,Zhang2014Quantum,PhysRevLett.114.233602},  authors have investigated the quantum speed limit  for cavity QED systems   and found that  the non-Markovianity  can speed up the quantum evolution. Some works have provided the quantum speed limit of  a central spin trapped in a spin-chain  environment to study the  behaviors of QSLT in the critical vicinity \cite{wei2016quantum,hou2015quantum}. The quantum speed limit in spin-boson models have also been studied in Refs.  \cite{Zhang2014Quantum,dehdashti2015decoherence}. Furthermore, to realize the controllable quantum evolution speed in open quantum systems, some methods have been proposed, such as  dynamical decoupling  pulses \cite{Song2017Control}, external classical driving field \cite{PhysRevA.91.032112} and optimal control \cite{caneva2009optimal}.  Most of these existing studies have been restricted to the environments with zero temperature, which motivates us  to do some investigations about the quantum speed limit in a finite-temperature environment. Besides, how to realize the controllable quantum  evolution speed in  a finite temperature environment is also one of issues that draw our attention.

In this paper, we firstly investigate the  quantum evolution speed of a  qubit which is locally coupled to its finite-temperature environment with Ohmic-like spectrum by using  the  stochastic decoupling method \cite{Shao2004,Shao2010Dissipative,PhysRevE.84.051112,Wu2014Double}. It is shown that the  quantum evolution speed of a qubit can be accelerated by the high  temperature in the strong-coupling regime. For the weak-coupling case, the bath temperature plays a role of dual character in affecting the  quantum evolution speed, which means that the high  temperature not only leads to the speed-up but also  speed-down processes. Furthermore, we find that the quantum evolution speed can be controlled by applying the  bang-bang pulse,  and  a relative steady value of quantum evolution speed is obtained. Interestingly, the bath temperature and  Ohmicity parameter also play roles of dual character in the strong-coupling regime since the presence of bang-bang pulse, which are not found in the case without pulse.
Secondly, we study the quantum evolution speed of a bare qubit coupled to a nonlinear thermal bath (spin-boson model) \cite{PRL.120401,Huang2010Effect} by applying the
 hierarchical equations of motion (HEOM) \cite{JPSJ.58.101,JPSJ.75.082001,PhysRevE.75.031107,1.2713104,1.3213013,PhysRevA.85.062323,1367-2630} which  is an effective numerical method that  beyond the Born-Markov approximation.
 It is  found that
the  quantum  evolution speed in  strong-coupling regime with low temperature
behaves similarly to that in the weak-coupling regime where the  quantum evolution speed can be accelerated  by the increase  of temperature. However, the rise of temperature induces the speed-down process in the strong-coupling regime with high temperature. As a comparison, the dynamics of quantum coherence is also explored in different situations.

 This paper is organized as follows. In Sec. \ref{sec2}, we investigate the quantum evolution speed of a qubit in bosonic environment by making use of the stochastic decoupling approach. In Sec. \ref{sec3}, we study the quantum evolution speed of a qubit in nonlinear environment by resorting to the HEOM method. The conclusion of this paper is given in Sec. \ref{sec4}.

\section{Quantum evolution speed in the bosonic environment}\label{sec2}

In this section, we study the quantum   evolution speed of a  qubit  coupled to its own finite-temperature environment.
First, we briefly outline the definition of quantum speed limit for an open quantum system.
 The maximal rate of evolution can  be characterized by the QSLT which is defined as the minimal time a quantum system needs to evolve from an initial state to a final state.
In open quantum systems,  the dynamical evolutions are governed  by the time-dependent master equation $L_t \rho_t = \dot{\rho}_t$ with $L_t$ being the positive generator of the dynamical semigroup.  Based on the relative purity along with von Neumann trace inequality and the Cauchy-Schwarz inequality, a unified lower bound on the QSLT including both MT and ML types  has been derived for arbitrary initial mixed states in the open quantum systems \cite{Zhang2014Quantum}, which reads
\begin{equation}
\tau_{\rm{QSL}}=max\left\lbrace\frac{1}{\overline{\sum_{i=1}^{n} \sigma_i \rho_i}},\frac{1}{\overline{\sqrt{\sum_{i=1}^{n} \sigma_i^2}}} \right\rbrace\ast\left| f_{ t +\tau_{\rm{D}}} -1 \right| Tr (\rho_t^2)
\label{qslt}
\end{equation}
where  $\overline{X} = \tau_{\rm{D}}^{-1} \int_{t}^{t+\tau_{\rm{D}}} X(t') \rm{d} t'$. $\sigma_i$ and $\rho_i$ are the singular values of $L_t \rho_t$ and $\rho_t$, respectively. $\tau_{\rm{D}}$ denotes the driving time.   The relative purity  between initial state $\rho_{t}$ and final state $\rho_{t+\tau_{\rm{D}}}$ of the quantum system is defined as $f_{t+\tau_{\rm{D}}} = Tr\left[\rho_{t+\tau_{\rm{D}}}\rho_{t}\right]/Tr(\rho^2_{t})$.

The first system under our consideration is  a  qubit  locally coupled to a finite-temperature bosonic  environment, also known as the spin-boson model, whose total Hamiltonian is described as ($\hbar=1$) \cite{Palma567}

\begin{equation}
\mathcal{H}= \frac{\Omega}{2} \sigma_z + \sum_{k} \omega_k b_k^\dagger b_k + \sum_{k} g_k \sigma_z  ( b_k^\dagger + b_k)
\end{equation}
where $\sigma_z$ is the standard Pauli operator in the $z$ direction, $b_k^\dagger$ and  $b_k$ denote the creation and annihilation operators of $k$th
oscillators with the frequency $\omega_k$, respectively.  The $g_k$  represents the coupling strength of qubit  to the finite-temperature   bath represented by a set of harmonic oscillators.
We investigate the dissipative quantum dynamics of the system by making use of the stochastic decoupling approach \cite{Shao2004,Shao2010Dissipative,PhysRevE.84.051112,Wu2014Double}, which is previously used in the calculation of partition functions and real-time dynamics for many-body systems.   Based on the Hubbard–Stratonovich transformation,  the dissipative interaction between the qubit  and the heat bath is decoupled via stochastic fields,  then the separated system and bath thus evolve in common  white noise fields.   The reduced density matrix comes out as an ensemble average of its random realizations. By resorting to the
Girsanov transformation \cite{Shao2004,PhysRevE.84.051112}, a  stochastic differential equation for
the random density matrix is obtained and can be used to  derive the desired master equation. Applying to the  system of  interest,  the master equation is given as
\begin{equation}
\label{master equation}
\frac{\rm{d}}{\rm{d} t} \rho_{\rm{s}} (t)= -\rm{i} \frac{\Omega}{2}\left[ \sigma_z, \rho_{\rm{s}} (t)\right] -\mathcal{D}(t)\left[\sigma_z, \left[\sigma_z, \rho_{\rm{s}} (t)\right]\right],
\end{equation}
where $\mathcal{D}(t)=\int_{0}^{t}\rm{d} t'C_{\rm{R}}(t')$ with $C_{\rm{R}} (t)$ being the real part of the correlation function $C(t)$. Assuming the bath is in a thermal equilibrium state $\rho_{\rm{b}}(0)=e^{-H_{\rm{b}}T^{-1}}/Tr_{\rm{b}}(e^{-H_{\rm{b}}T^{-1}})$ with the Boltzmann constant $k_{B}=1$, then the correlation function for this bosonic bath is given by
\begin{equation}
\label{correlation function}
C(t)= \int \rm{d}\omega J(\omega) \left[ \coth \left(\frac{\omega}{2T}\right) \cos(\omega t) -\rm{i}\sin(\omega t) \right],
\end{equation}
 in which $J(\omega)$ is the  bath spectral density function, and $T$ represents the temperature. As a result,  the reduced density matrix of qubit can be obtained by solving  the Eq. (\ref{master equation}):
 \begin{equation}
\label{reduced matrix A}
{\rho _{\rm{s}}}(t) = \left( {\begin{array}{*{20}{c}}
	{{\rho_{\rm{ee}}}(0)}&{{\rho_{\rm{eg}}}(0){e^{\rm{i}{\Omega}t - \Gamma (t)}}} \\
	{{\rho_{\rm{ge}}}(0){e^{ - \rm{i}{\Omega}t - \Gamma (t)}}}&{{\rho_{\rm{gg}}}(0)}
	\end{array}} \right),
\end{equation}
where
\begin{equation}
\label{decoherence factor}
\Gamma(t)= 4\int_{0}^{\infty} \rm{d}\omega J(\omega) \frac{1-\cos(\omega t)}{\omega^2} \coth \left(\frac{\omega}{2T}\right)
\end{equation}
 is the decoherence factor.  In this article, we consider  the spectral density of the environmental modes is Ohmic-like $J(\omega)=\Lambda (\omega^s/\omega^{s-1}_{\rm{c}}) e^{-\omega/\omega_{\rm{c}}}$ with $\Lambda$ being the dimensionless coupling constant and $\omega_{\rm{c}}$ being the cutoff frequency. It is possible to obtain Ohmic reservoirs ($s=1$) and sub-Ohmic reservoirs($s<1$)  by changing the Ohmicity parameter $s$.

In the Bloch sphere representation, a generally mixed state $\varrho$ of a qubit can be written in terms of Pauli matrices $\varrho = \frac{1}{2} (I + v_x \sigma_x +v_y \sigma_y + v_z \sigma_z)$, where the coefficients $v_x$, $v_y$, $v_z$ are the Bloch vector, and $I$ is the identity operator of the qubit. The time evolution of the reduced density matrix $\rho_t$ in the Bloch sphere representation can be derived from  Eq. (\ref{reduced matrix A})
\begin{equation}
{\rho _{\rm{s}}}(t) =\frac{1}{2}\left( {\begin{array}{*{20}{c}}
	{1 + {v_z}}&{({v_x} - \rm{i}{v_y}){q_t}} \\
	{({v_x} + \rm{i}{v_y}){q^{*}_t}}&{1 - {v_z}}
	\end{array}} \right),
\end{equation}
where $q_t = e^{\rm{i}{\Omega}t - \Gamma_t}$ with $\Gamma_t$ being the decoherence factor, see Eq. (\ref{decoherence factor}). It is readily find that the excited state population is unchanged, thus the evolution of qubit  is a dephasing process. According to Eq. (\ref{qslt}),  the QSLT for the qubit  to evolve from initial state $\rho_t$ to final state $\rho_{t +\tau_{\rm{D}}}$ in this  dephasing model is given by
 \begin{equation}
\label{qsltcoherence}
\tau_{\rm{QSL}} =\frac{\frac{1}{2} \sqrt{v_x^2 +v_y^2} \left|(q_t-q_{t+\tau_{\rm{D}}})q_t^{*} +H.c. \right|}{\frac{1}{\tau_{\rm{D}}} \int_{t}^{t+\tau_{\rm{D}}} \left|\dot{q}_{t'} \right| \rm{d} t'}.
\end{equation}
 \begin{figure}[ht]
	\centering
	\subfigure{ \label{figure1a} \includegraphics[width=8cm]{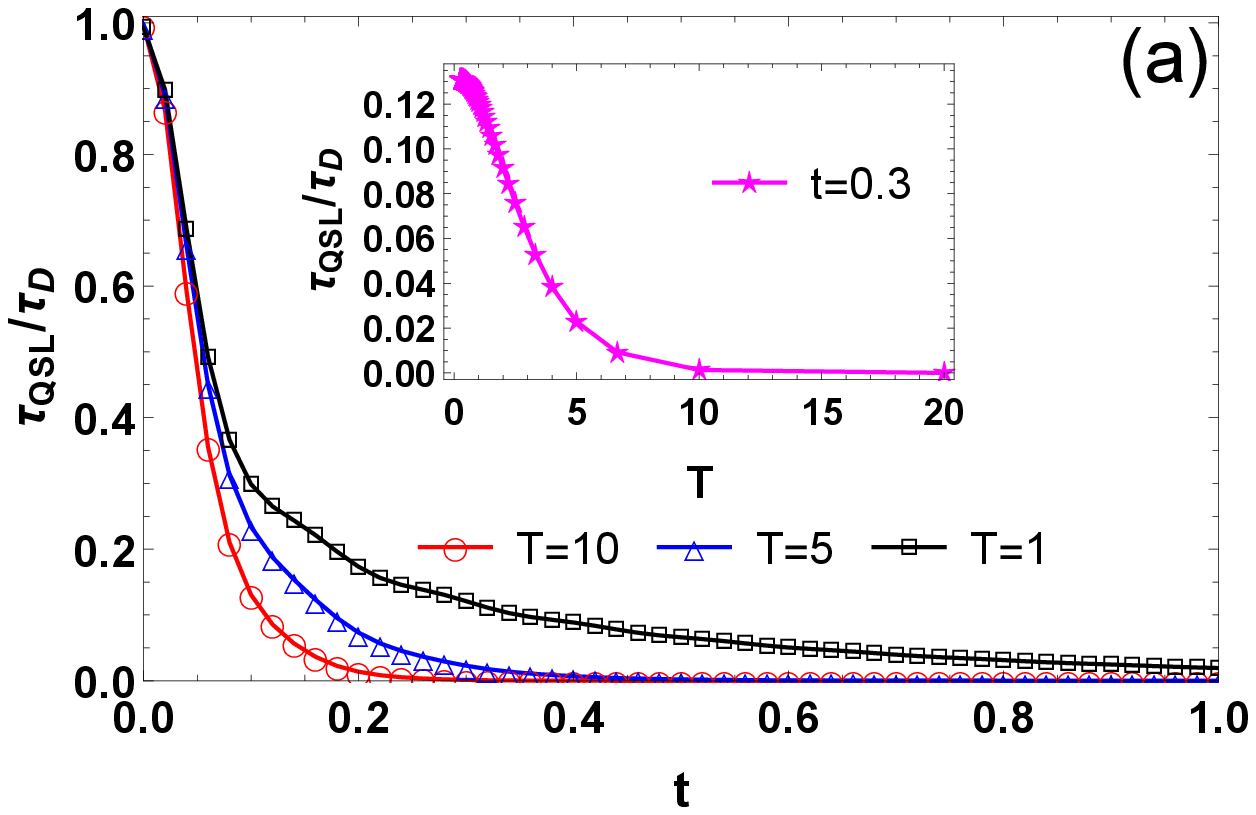}}
	\subfigure{\label{figure1b}
		\includegraphics[width=8cm]{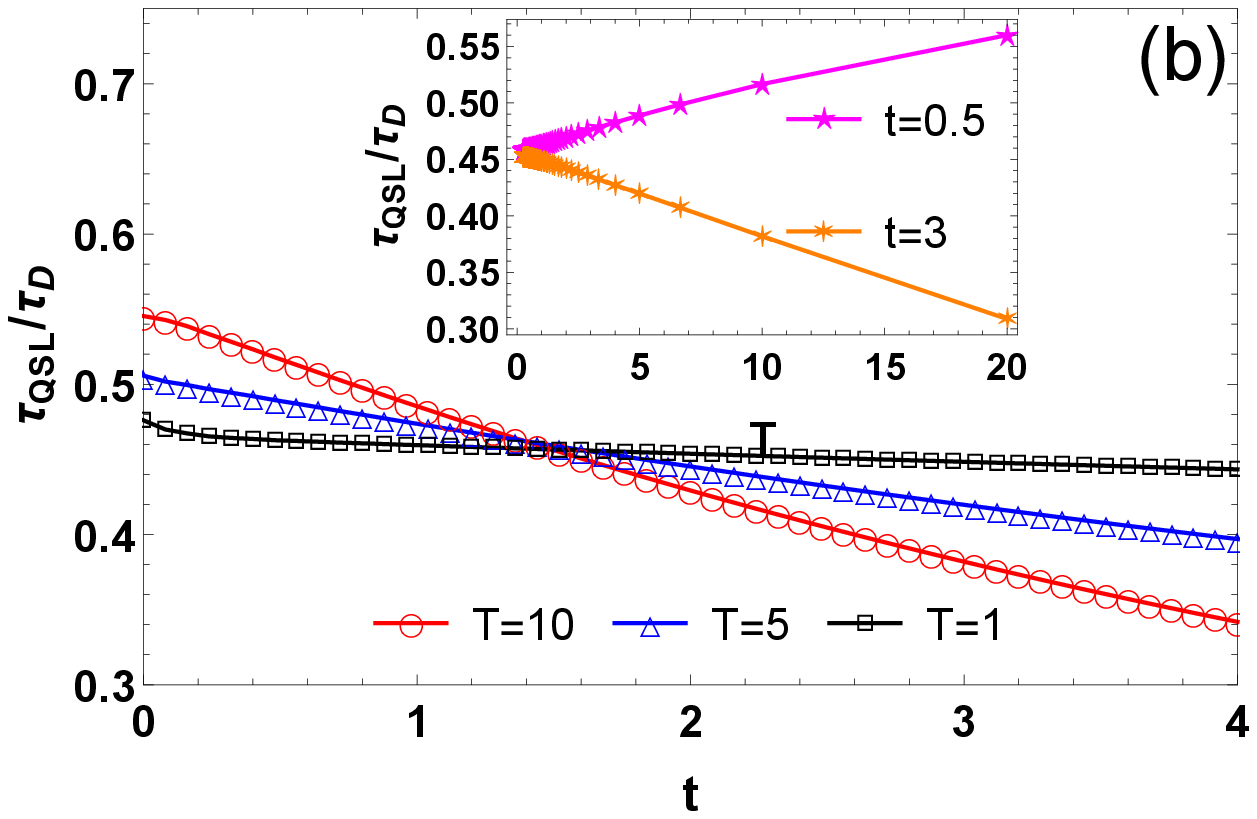}}
	\caption{(a) The QSLT $\tau_{\rm{QSL}}/\tau_{\rm{D}}$    as a function of the  initial time  parameter $t$ for different temperatures  in the strong-coupling regime $\Lambda=0.2$. The inset shows the QSLT $\tau_{\rm{QSL}}/\tau_{\rm{D}}$  as a function of the temperature $T$ for $t=0.3$.
		(b) The QSLT $\tau_{\rm{QSL}}/\tau_{\rm{D}}$  as a function of the  initial time  parameter $t$ for different temperatures  in the weak-coupling regime $\Lambda=0.001$.  The inset shows  the QSLT $\tau_{\rm{QSL}}/\tau_{\rm{D}}$  as a function of the temperature $T$ for  $t=0.5$ and $t=3$. Here we set the driving time $\tau_{\rm{D}}=1$. Other parameters are $\Omega=1$, $\omega_{\rm{c}}=50$ and $s=1$.}
\end{figure}

 %%%%%%%%%%%%%%%%%%%%%%%%%%%%%%%%%%%%%%%%%%%%%%%%%%
\subsection{The evolution of quantum speed limit}
In this section, we  give the results about the evolution of quantum speed limit for a qubit locally coupled to a finite-temperature  environment. For simplicity, we assume the initial state of qubit is $\rho_0=1/2(\left|0\right\rangle \left\langle 0\right|+\left|0\right\rangle \left\langle 1\right|+\left|1\right\rangle \left\langle 0\right|+\left|1\right\rangle \left\langle 1\right|)$  and fix the computational basis $\left\lbrace |0\rangle, |1\rangle \right\rbrace $ as the reference basis, in which $|0\rangle$ and $|1\rangle$ are the ground  and  excited states
of Pauli operator $\sigma_z$, respectively.
In Fig. \ref{figure1a}, we plot the QSLT  for the Ohmic dephasing model as a function of initial time   parameter $t$ with different temperatures $T$  in the strong-coupling  regime. The inset shows the variation of  QSLT as a function of temperature $T$ for a fixed initial time   parameter $t=0.3$. We can find that the  QSLT  of qubit  decays monotonically to zero  with the growth of time $t$, and the more the   temperature increases,  the smaller the QSLT becomes, which means that the high  temperature  speed up the   evolution of qubit. This result can be understood by the fact that a higher bath temperature  induces a more severe decoherence which leads to the acceleration of quantum evolution.  In the weak-coupling regime,   it can be seen from Fig. \ref{figure1b} that the temperature shows different  effects on the QSLT  compared to what it exhibits in the strong-coupling regime. The high  temperature initially prolongs the QSLT and suppresses the  quantum  evolution speed, however, some time later, the high  temperature can speed up the quantum evolution, which means that the bath temperature exhibits two  sides for the quantum  evolution  speed  in the weak-coupling regime. When the temperature increases, the decay rate of QSLT is enhanced. Moreover, we can observe from the inset of Fig. \ref{figure1b} that  a asymptotic value of QSLT below the driving time is obtained  as the zero temperature is approached. This result suggests that the extreme low bath temperature may freeze the speed of evolution of qubit  in this dissipative system.

\begin{figure}[!ht]
	\centering
	\subfigure{ \label{figure2a} \includegraphics[width=9cm]{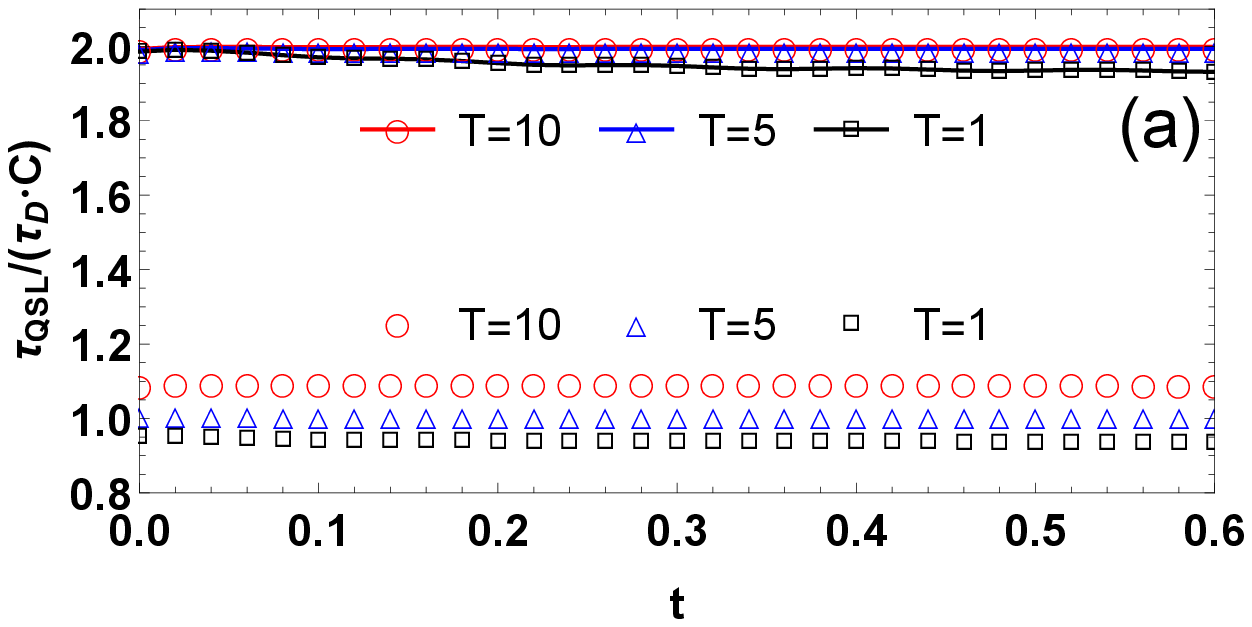}}
	\subfigure{ \label{figure2b} \includegraphics[width=8cm]{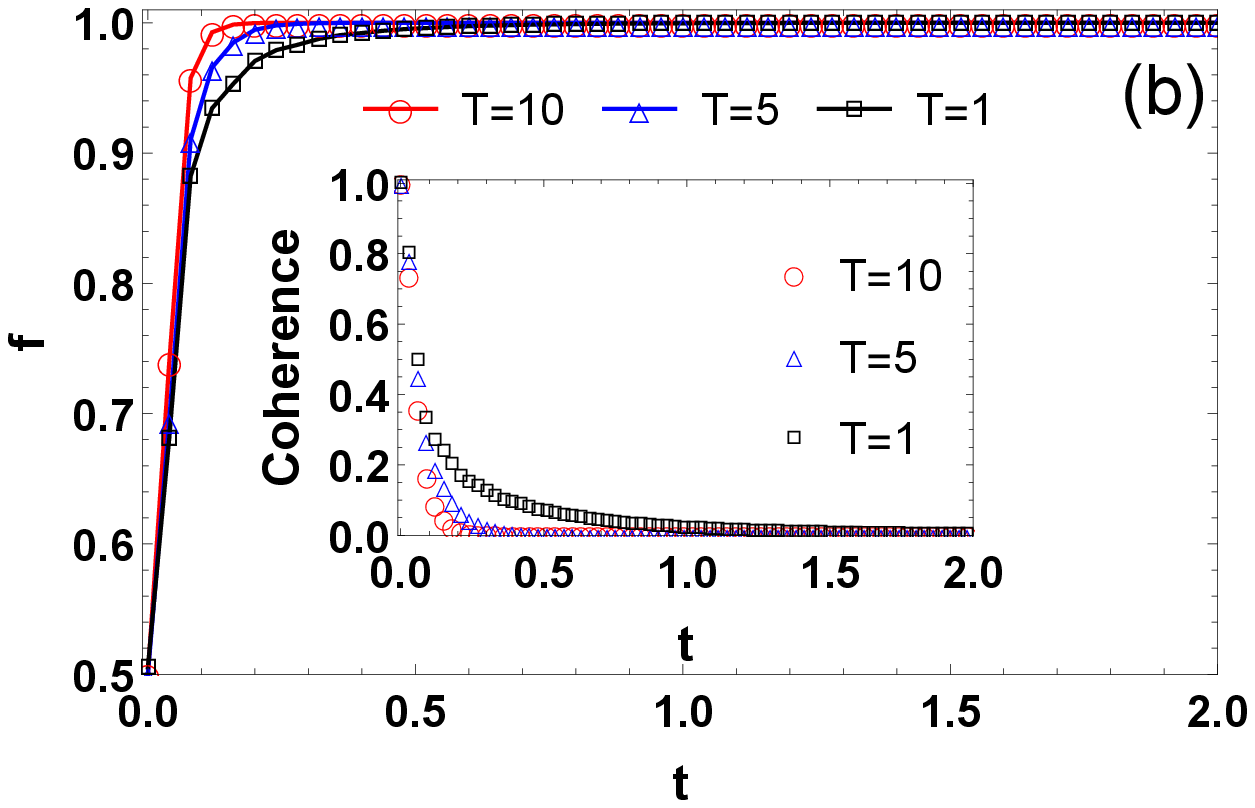}}
	\subfigure{\label{figure2c} \includegraphics[width=8cm]{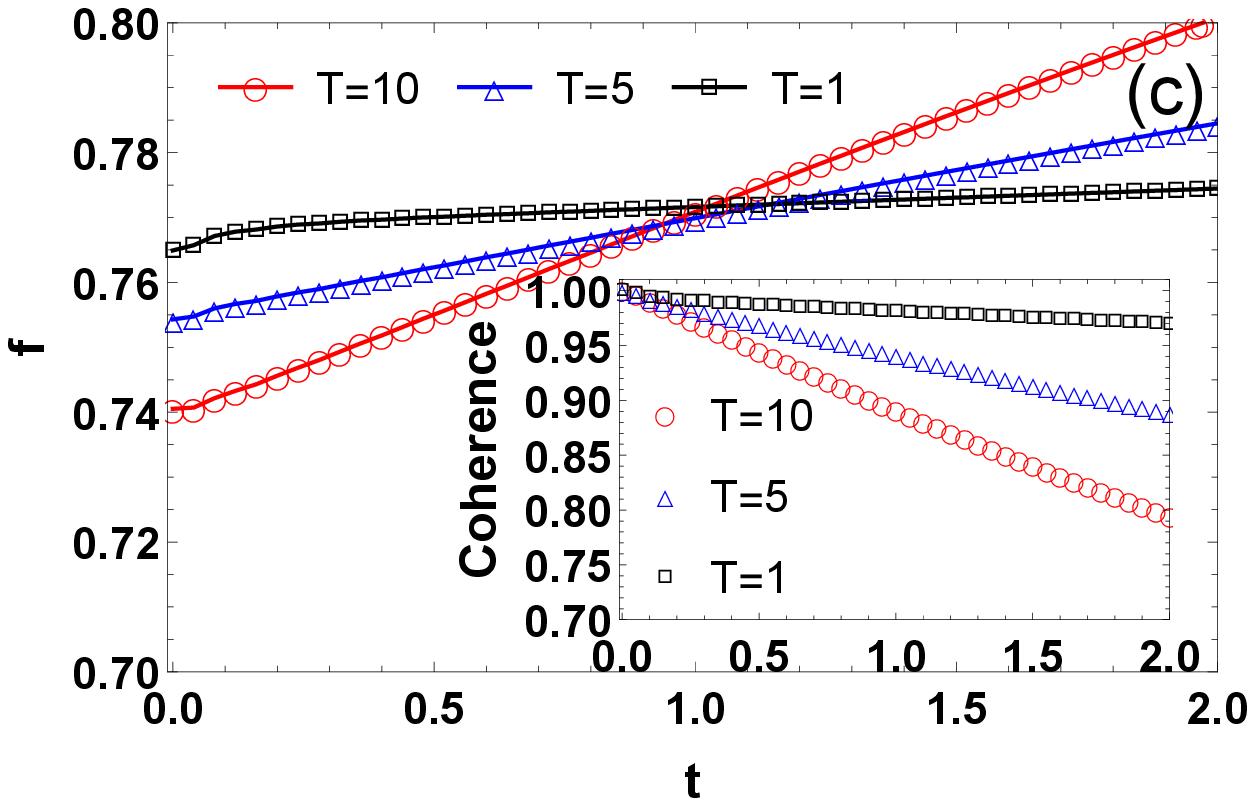}}
	\caption{(a) The ratio of QSLT to the $l_1$-norm quantum coherence $\tau_{\rm{QSL}}/(\tau_{\rm{D}}\cdot C)$ as a function of the  initial time  parameter $t$ for different temperatures  in the weak-coupling regime $\Lambda=0.001$  (dashed line)and  strong-coupling regime $\Lambda=0.2$ (solid line). The relative purity $f$   as a function of the  initial time  parameter $t$ for different temperatures  in the (b) strong-coupling regime $\Lambda=0.2$ and  (c) weak-coupling regime $\Lambda=0.001$.  The insets show the time evolutions of $l_1$-norm coherence for different temperatures in the (b) strong-coupling regime $\Lambda=0.2$ and  (c) weak-coupling regime $\Lambda=0.001$. Here we set the driving time $\tau_{\rm{D}}=1$. Other parameters are $\Omega=1$, $\omega_{\rm{c}}=50$ and $s=1$.}
\end{figure}
Here, we would like to provide a possible physical explanation why the QSLT shows different behaviors in the strong-coupling and weak-coupling regimes for changing temperatures in this quantum dissipative system. In the dephasing model,  it has been found that the QSLT  relates to the quantum coherence of the initial state  under a given driving time $\tau_{\rm{D}}$ \cite{Zhang2014Quantum}, which can also be confirmed from the  term $ \sqrt{v_x^2 +v_y^2} $ in Eq. (\ref{qsltcoherence}). If we do a further derivation,  Eq. (\ref{qsltcoherence}) can be rewritten as
\begin{equation}
\label{qsltcoherencet}
\tau_{\rm{QSL}} = C_t \cdot\frac{{\left| {{e^{ - {\Gamma _t}}} - \cos (\Omega {\tau_{\rm{D}}}){e^{ - {\Gamma _{t + {\tau_{\rm{D}}}}}}}} \right|}}{{\frac{1}{{{\tau_{\rm{D}}}}}\int_t^{t + {\tau_{\rm{D}}}} {{e^{ - {\Gamma _{t'}}}}\sqrt {(\Gamma _{t'}^2 + {\Omega ^2}{{t'}^2})(\dot \Gamma _{t'}^2 + {\Omega ^2})} \rm{d} t'} }}
\end{equation}
 in which $C_t = \sqrt{v_x^2 +v_y^2} e^{-\Gamma_{t}}$ is the $l_1$-norm quantum coherence of qubit  according  to the definition in Ref. \cite{PhysRevLett.Coherence}. Based on Eq. (\ref{qsltcoherencet}), we can  explore the relationship between the QSLT and quantum coherence in  this dephasing model. It is clear that the QSLT at time $t$ is directly related to the quantum coherence of the same time. To get more insight on the role of temperature in the quantum evolution speed, we utilize the   ratio of  QSLT  to quantum coherence   to analysis the different phenomenons observed above.  The ratio as a function of the initial time parameter for  different temperatures  is  displayed in Fig. \ref{figure2a}. We can observe that in the strong-coupling regime, the ratios have little gaps at first for different temperatures, and then the gaps enlarge gradually as time goes on, whereas   the ratio  always have relatively stable gaps from beginning  for different temperatures in the weak-coupling regime.  This result suggests that the second term on the right of Eq. (\ref{qsltcoherencet}) may lead to the different performances of quantum evolution speed, although which  only gives a superficial interpretation, it provides inspiration for further studying.

 Since the  expression of QSLT in Eq. (\ref{qslt}) is based on the relative purity,  we mainly focus on the investigation about relative purity in the following. In Figs.\ref{figure2b} and \ref{figure2c},   we plot the time evolutions of relative purity  and $l_1$-norm coherence for different temperatures in the strong-coupling and weak-coupling regimes, respectively. The relative purity gradually increases to the maximum as time goes on  in the strong-coupling regime, and the higher temperature leads to faster increase. Instead, the quantum coherence gradually decreases to the minimum  in the time evolution, and the higher temperature  leads to faster decrease. This is due to the fact that the increase of temperature  brings about more intense thermal fluctuation which induces the stronger decoherence.
Similar behaviors can be found in the weak-coupling case which is displayed in Fig. \ref{figure2c}.  One difference is that the changing rates  of relative purity and quantum coherence are smaller than the ones in the strong-coupling case. Another difference is that the values of relative purity at the initial time $t=0$ are not consistent which leads to the   curves for different temperatures cross each other.
Comparing to the Fig. \ref{figure1b}, we can find that the  bath temperature plays a role of dual character in affecting the QSLT and this phenomenon may be linked to  the performance of relative purity.  It is clearly observed from the time evolutions of quantum coherence that the finial states $\rho_{0+\tau_{\rm{D}}}$ with driving time $\tau_{\rm{D}}=1$ under various temperatures  have little difference in the strong-coupling regime, while have obvious gaps in the weak-coupling case, which contributes to the initial values of relative purity are consistent in the strong-coupling regime, however, are  different in the weak-coupling case.

Thus,  the reason why the quantum evolution speed shows different performances in the  strong-coupling and weak-coupling  regimes is threefold: first, the quantum evolution speed  at time $t$ is related to the quantum coherence at the same moment. Second,  the higher temperature  brings about more intense thermal fluctuation in the strong-coupling regime than in the weak-coupling case, which leads to the  quantum coherence decreasing faster in the strong-coupling regime than in the weak-coupling  case. Third, the initial values of relative purity depends on the driving time which may contribute to the different behaviors of relative purity for various temperatures  in strong-coupling and weak-coupling regimes. Above discussions may help us understand  the effects of bath temperature on the quantum evolution speed of this dephasing model, and realize that the environment-assisted speed-up and speed-down processes are possible.

\begin{figure}[htbp]
	\centering
		\subfigure{ \label{figure3a}
		\includegraphics[width=6cm]{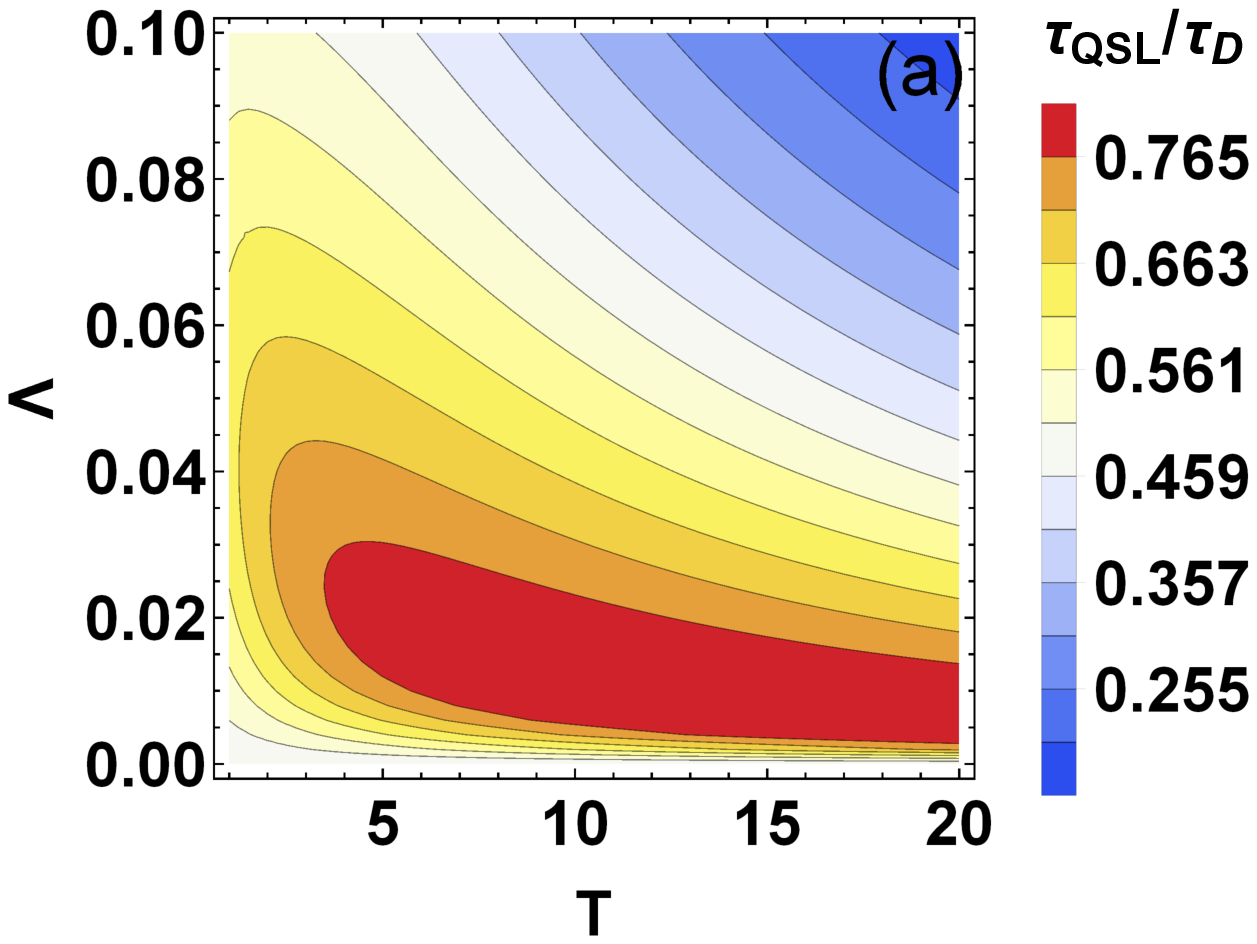}}
	\subfigure{ \label{figure3b}
		\includegraphics[width=6cm]{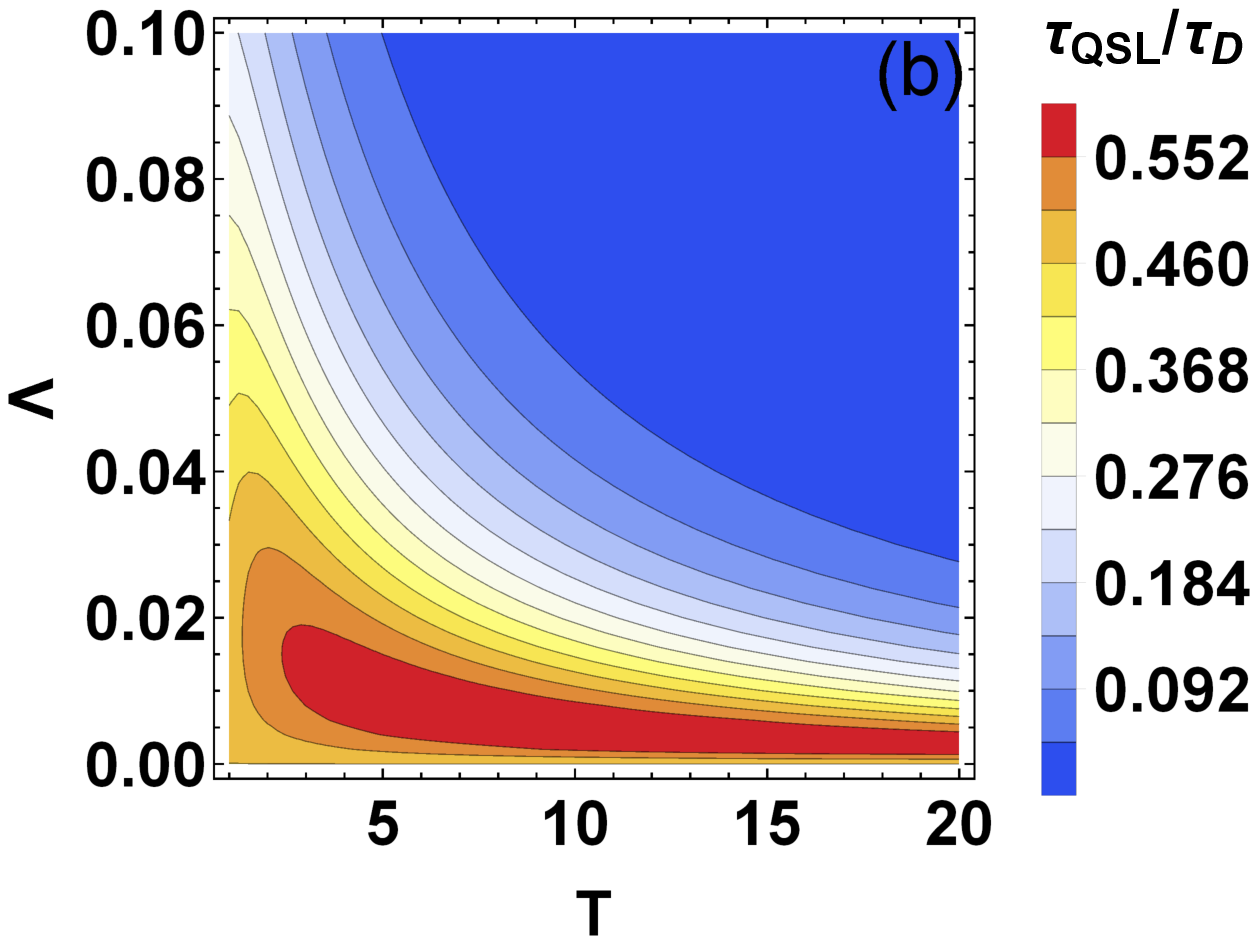}}
	\subfigure{ \label{figure3c}
		\includegraphics[width=6cm]{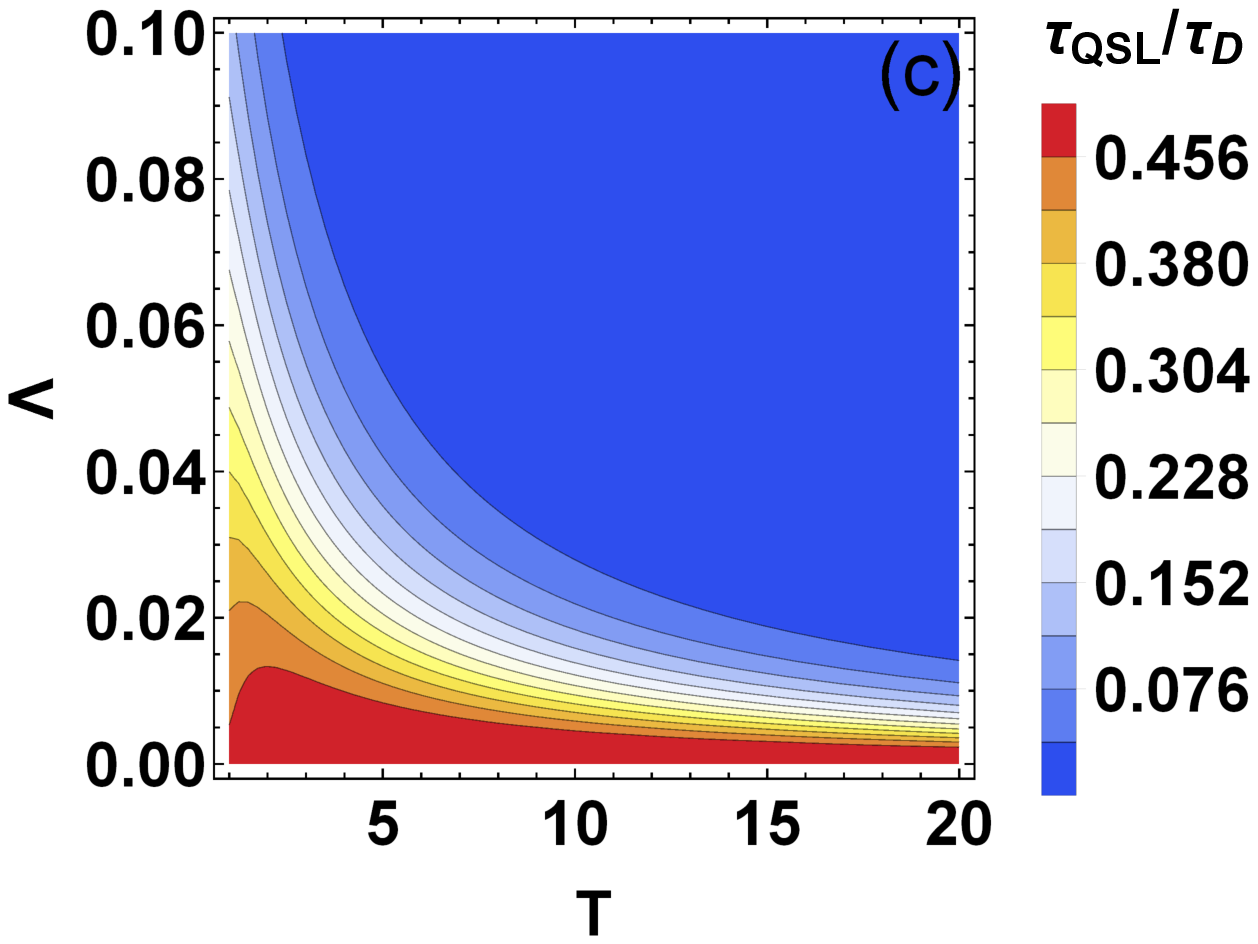}}
	\subfigure{ \label{figure3d}
		\includegraphics[width=6cm]{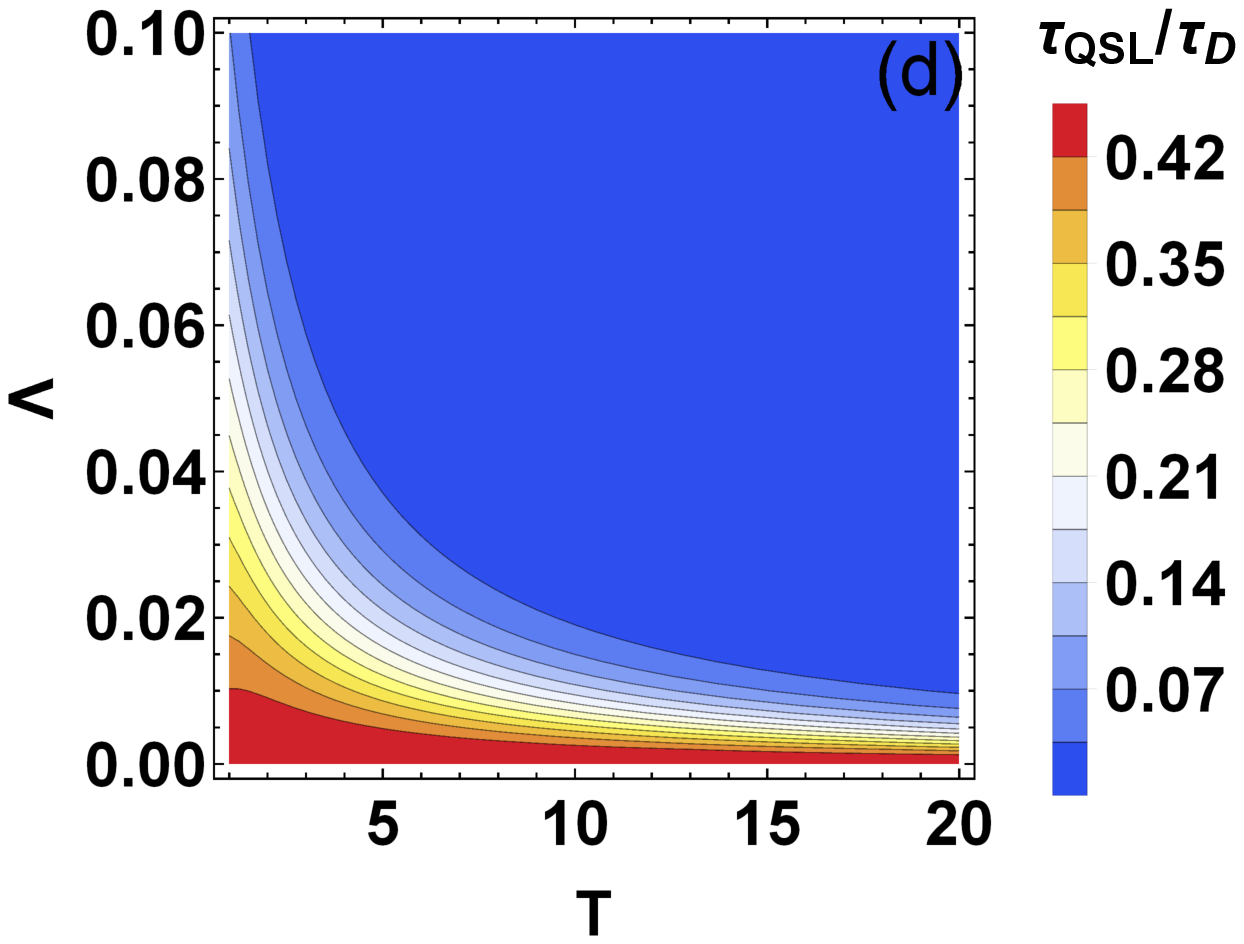}}
	\caption{  Contour plot of variation of the QSLT $\tau_{\rm{QSL}}/\tau_{\rm{D}}$ as a function of temperature $T$ and coupling strength $\Lambda$ for (a) $t=0.1$, (b) $t=0.5$, (c) $t=1$ and (d) $t=1.5$. Other parameters are  $\tau_{\rm{D}}=1$, $\Omega=1$, $\omega_{\rm{c}}=50$ and $s=1$. }
	\label{figure3}
\end{figure}

  We display the contour plot of QSLT  as a function of bath temperature $T$ and coupling strength $\Lambda$ for different  initial  time parameters in  Fig. \ref{figure3}. It is quite clear from  Fig. \ref{figure3}(a) that the QSLT has a peak in the weak-coupling regime with the initial time parameter $t=0.1$, which means that the QSLT  increases at first, then decreases along with the growth of bath temperature. The bath temperature plays a role of  dual character  in affecting the quantum evolution speed  in the weak-coupling regime. By contrast, in the strong-coupling regime, the QSLT  only decreases with the increase of the bath temperature.   Furthermore, it is clearly observed from the Fig. \ref{figure3}(b)-(d) that the dual character of temperature exhibited in the weak-coupling regime gradually disappears as the time goes on, and then  the quantum  evolution speed is accelerated by increasing temperature in both strong-coupling and weak-coupling regimes. Therefore, the bath temperature  plays a more important and broader role in the weak-coupling regime than it does in the strong-coupling one.

\begin{figure}[t]
	\centering
	\subfigure[]{ \label{figure4a}
		\includegraphics[width=7cm]{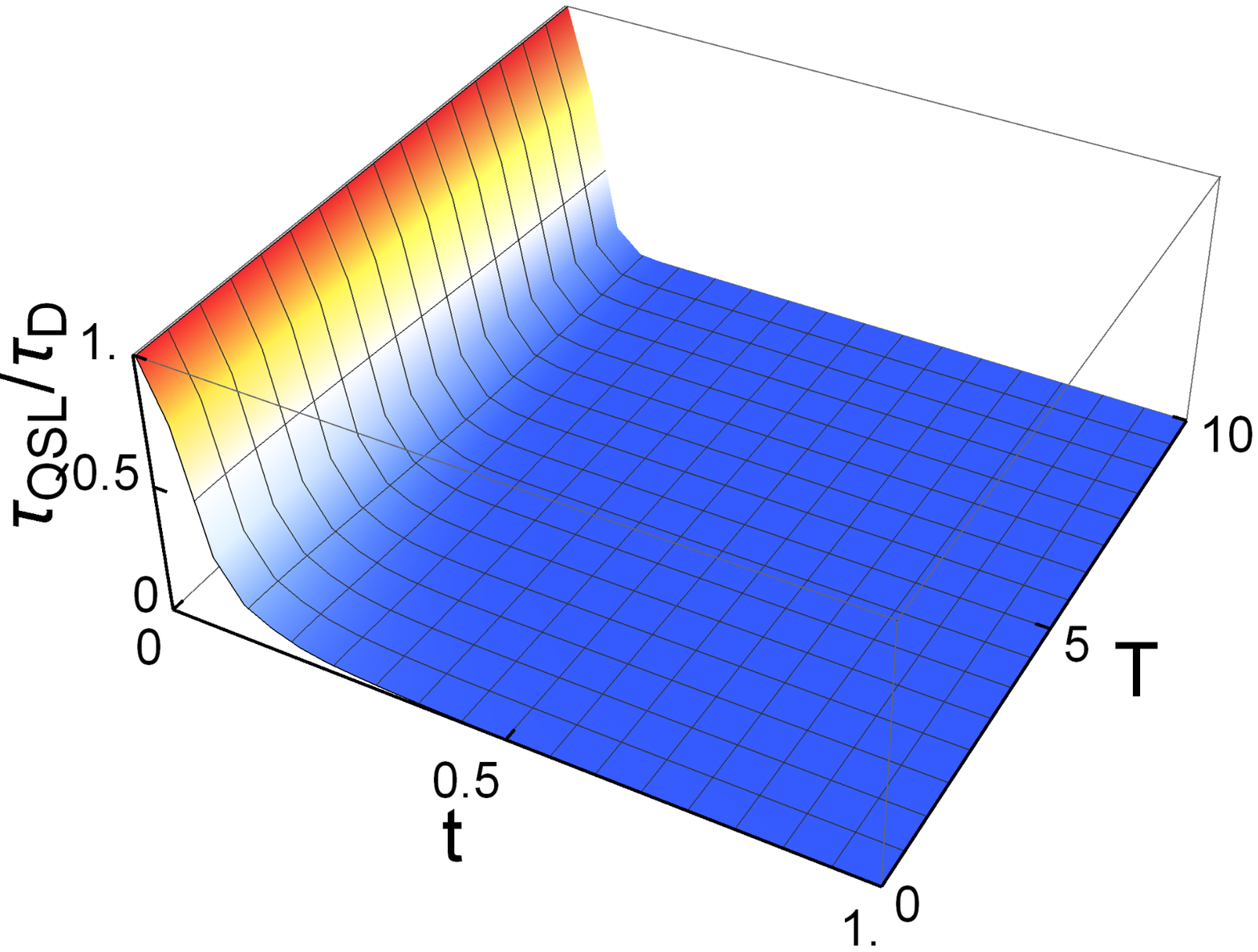}}
	\subfigure[]{ \label{figure4b}
		\includegraphics[width=7cm]{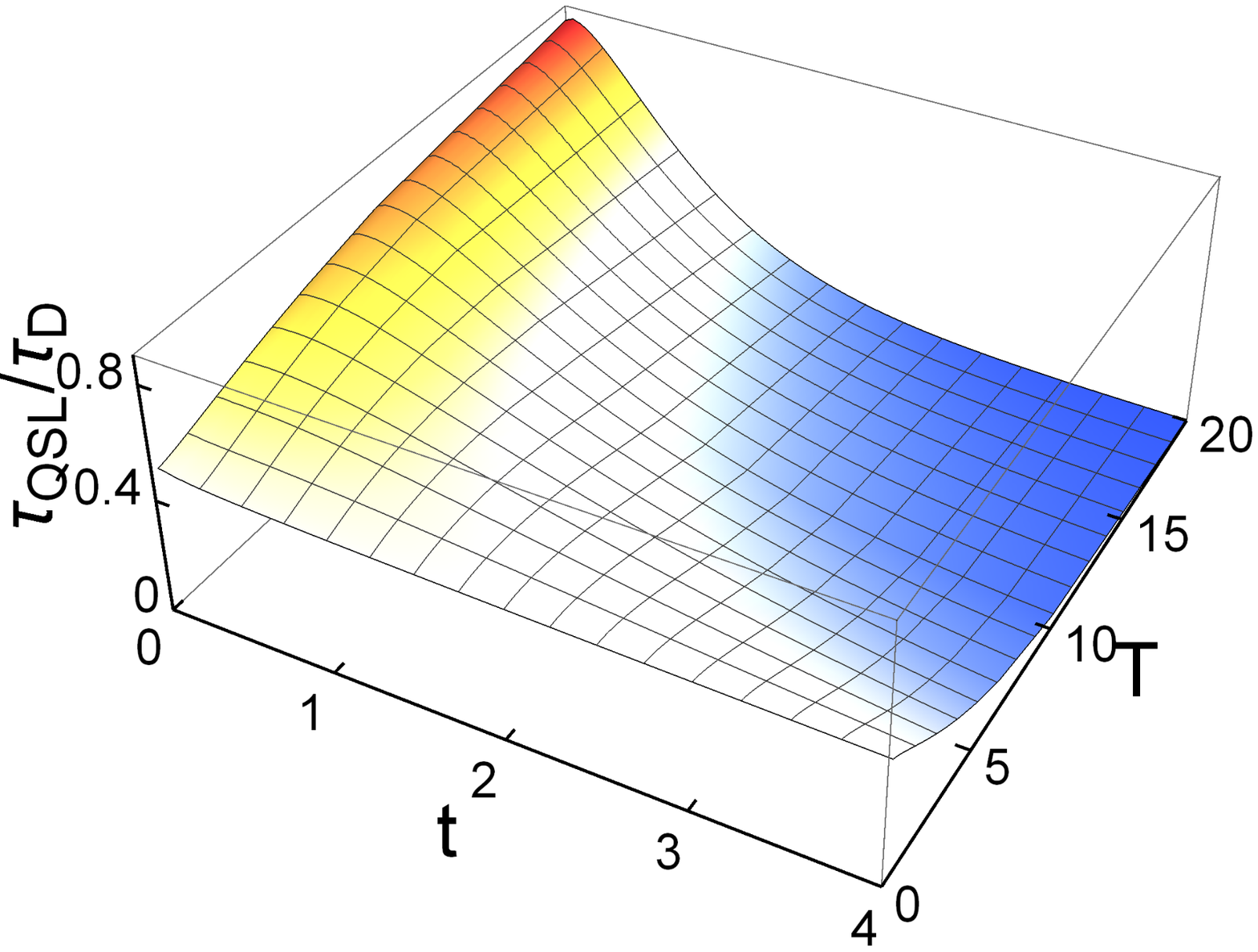}}
	\subfigure[]{ \label{figure4c}
		\includegraphics[width=7cm]{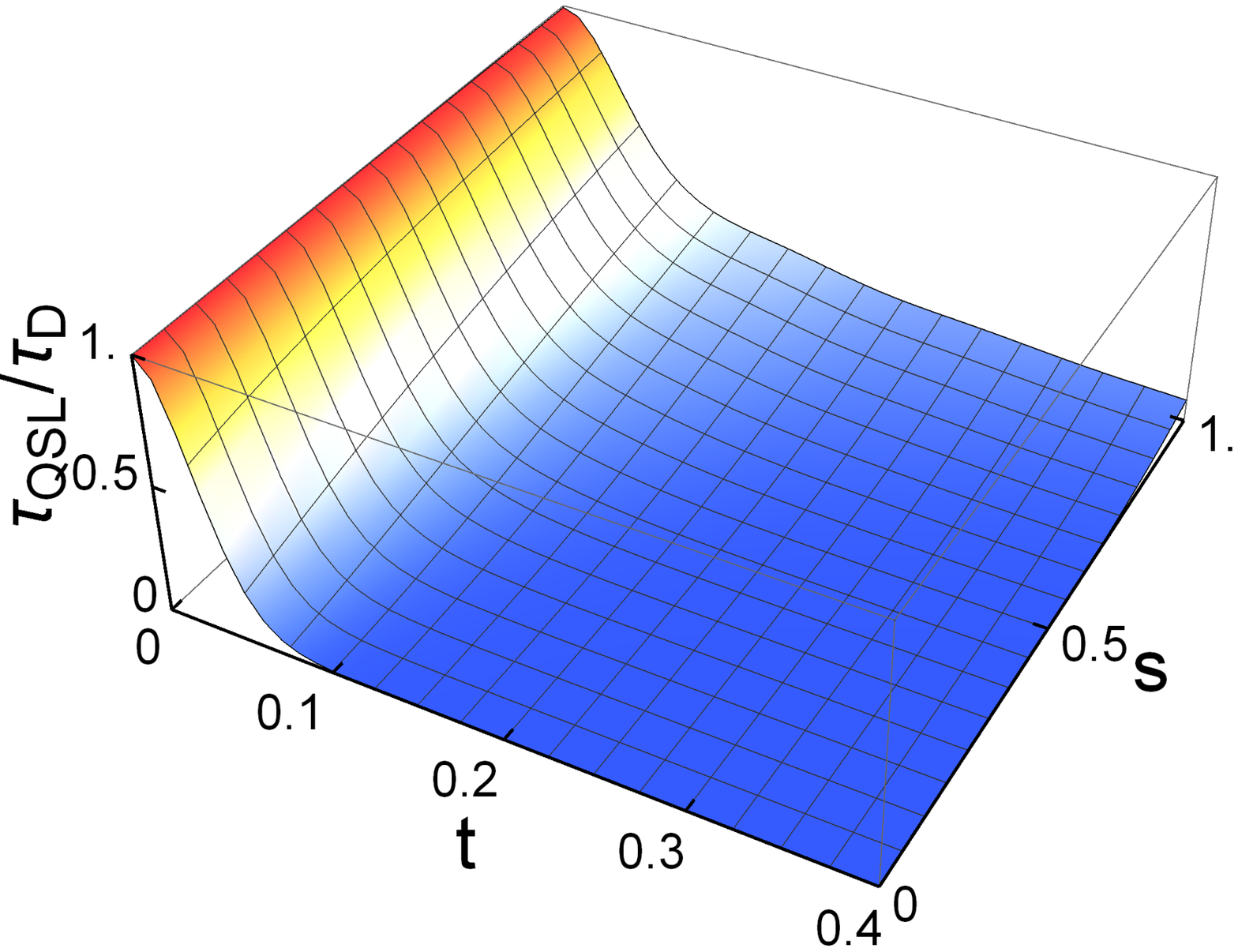}}
	\subfigure[]{ \label{figure4d}
		\includegraphics[width=7cm]{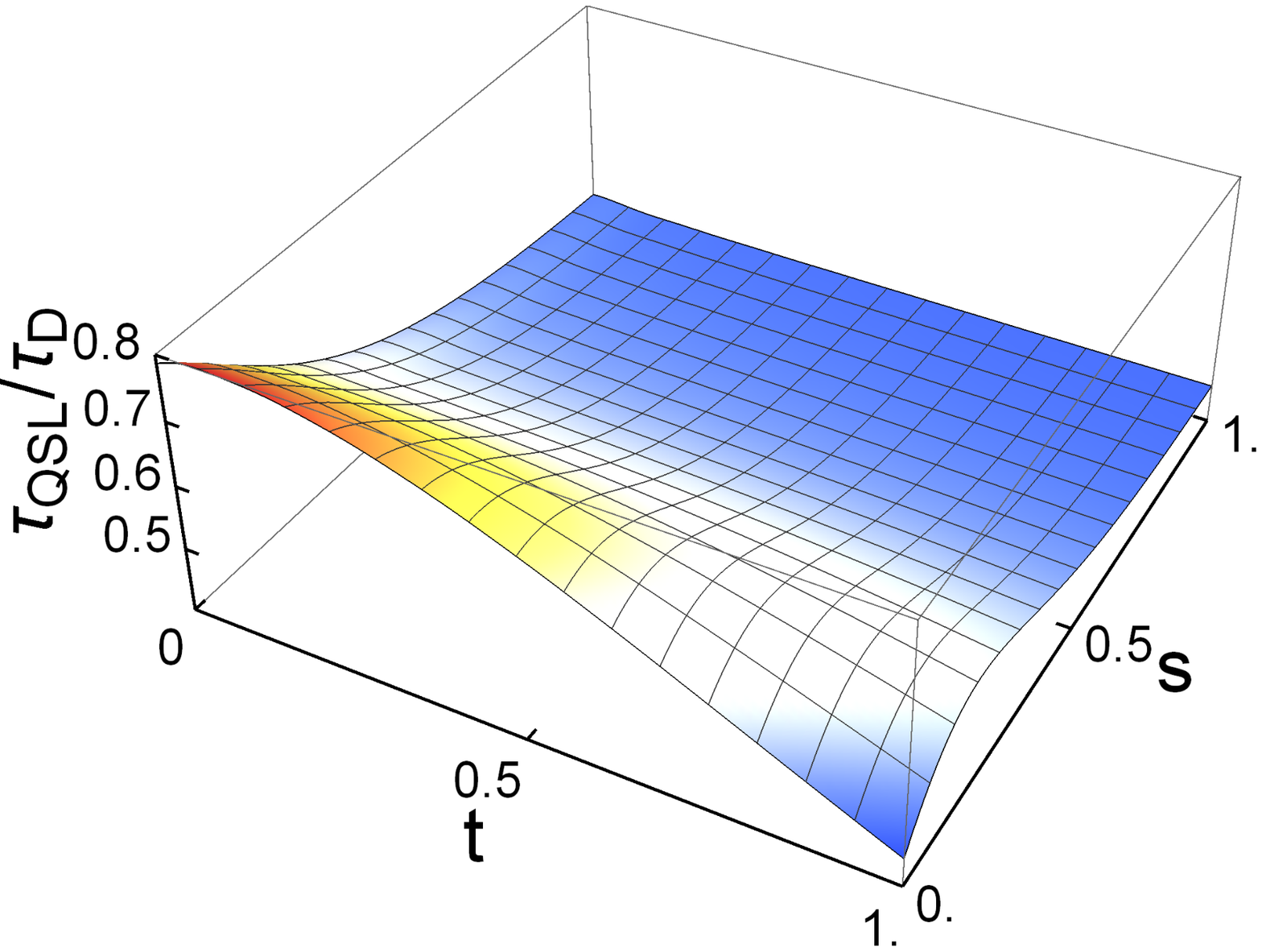}}
	\caption{The QSLT $\tau_{\rm{QSL}}/\tau_{\rm{D}}$ of qubit   as
		functions of the initial time  parameter $t$ and bath temperature $T$  for the sub-Ohmic
		spectrum $s=0.6$  in the (a) strong-coupling regime  $\Lambda=0.2$ and (b) weak coupling regime  $\Lambda=0.001$, respectively. The QSLT $\tau_{\rm{QSL}}/\tau_{\rm{D}}$ of qubit   as
		functions of the  Ohmicity parameter $s$ and initial time  parameter $t$ for bath temperature $T=1$ in the (c) strong-coupling regime $\Lambda=0.2$ and (d) weak-coupling regime $\Lambda=0.001$, respectively. Other Parameters are chosen as $\tau_{\rm{D}}=1$, $\omega_{\rm{c}}=50$ and $\Omega =1$.  }
\end{figure}

In the following, we explore the  variations of QSLT  of qubit  for the sub-Ohmic reservoirs with different relevant parameters. We plot the  QSLT    as
functions of the initial time  parameter $t$ and bath temperature $T$  for the sub-Ohmic  spectrum $s=0.6$  in the  strong-coupling regime  [Fig. \ref{figure4a}] and  weak coupling regime  [Fig. \ref{figure4b}], respectively. It is found that the effects of  bath temperature   on the  QSLT of qubit  for  the sub-Ohmic  spectrum are similar to  the case for Ohmic  spectrum. We can see from   Fig. \ref{figure4a} that  the  QSLT   decreases monotonically with the initial time  parameter $t$ in the  strong-coupling regime, and the increase of bath temperature leads to the shorter QSLT, namely, the faster quantum evolution.  In the weak-coupling regime,  as shown in Fig. \ref{figure4b},  the  bath temperature also plays a role of  dual character  in influencing the  speed of evolution.  Initially, the  QSLT   is a monotonic increasing function of bath temperature $T$, however, some time later, it is changed to be a monotonic decreasing function of  $T$. Moreover, a relative steady speed of evolution can be obtained at the zero temperature, which is a unique phenomenon in weak coupling regime.

To get more insight on the role of the sub-Ohmic   reservoirs  in influencing the QSLT, we display the QSLT as functions of the  Ohmicity parameter $s$ and initial time  parameter $t$ in the  strong-coupling regime  [Fig. \ref{figure4c}]and  weak-coupling regime  [Fig. \ref{figure4d}], respectively.  It is clear to see that the QSLT  is indeed  extended  with the increase of the  Ohmicity parameter $s$ in the strong-coupling regime and the QSLT  is not a simple monotone function of the Ohmicity parameter $s$ in the  weak-coupling regime.  In comparison, we can see from  Fig. \ref{figure4d} that the QSLT  decreases to a minimum with the growth of the Ohmicity parameter $s$  in the beginning of the evolution, and after a certain time  the  QSLT first increases to a maximum  with increasing $s$, and then decreases with further increase of $s$, which means that a nonmonotonic behavior of the  QSLT is displayed. Furthermore,  the quantum evolution  speed of qubit   for sub-Ohmic spectrum is faster than the one for Ohmic spectrum in the  strong-coupling regime, which is inverse in the  weak-coupling regime.

%%%%%%%%%%%%%%%%%%%%%%%%%%%%%%%%%%%%%%%%%%%%%%%%%%%%%%%%
\subsection{Control of the quantum evolution speed by applying bang-bang pulses}

\begin{figure}[htbp]
	\centering
	\includegraphics[width=9cm]{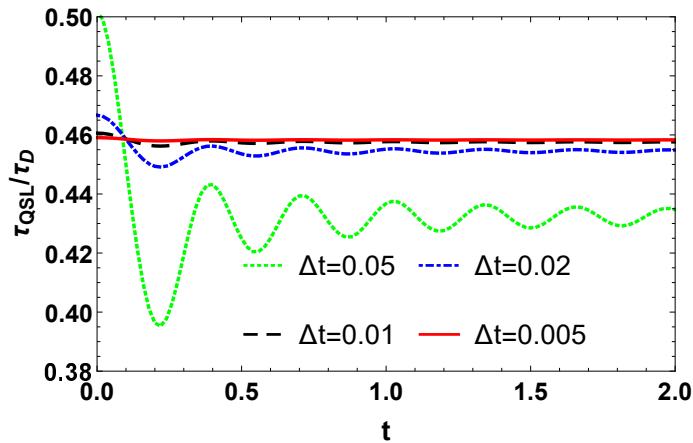}
	\caption{ The QSLT $\tau_{\rm{QSL}}/\tau_{\rm{D}}$ of qubit   versus the initial time  parameter $t$ with different  pulse interval : $\Delta t =0.05$ (green dotted line), $\Delta t =0.02$ (blue dotted-dashed line), $\Delta t =0.01$ (black dashed line), $\Delta t =0.005$ (red solid line) for the Ohmic spectrum  in the  strong-coupling regime $\Lambda=0.2$. Other parameters are $\tau_{\rm{D}}=1$, $\omega_{\rm{c}}=20$, $\Omega =1$ and $T=1$. }
	\label{figure5}
\end{figure}
The major obstacle to the development of quantum technologies is the destruction of all quantum properties caused by the inevitable interaction of quantum systems with their surrounding environment. Much effort has been made to minimize the influence of environmental noise or suppress the  decoherence induced by environment in the practical realization of the quantum tasks. One of the interesting approaches is the \textquotedblleft  dynamical decoupling\textquotedblright or \textquotedblleft bang-bang\textquotedblright pulses \cite{PhysRevAbangbang,PhysRevA.69.030302,PhysRevLett.82.2417}, which is based on applying the strong and sufficiently fast pulses to restore the quantum coherence of target system.

In this section, we mainly investigate the effect of bang-bang pulses on the quantum evolution speed of qubit. The Hamiltonian of control pulses is given by \cite{PhysRevA.69.030302}
\begin{equation}
H_{\rm{p}} (t) = \sum_{n=1}^{N} \mathcal{A}_n (t) e^{\rm{i}\Omega t\sigma_z/2} \sigma_x e^{-\rm{i}\Omega t\sigma_z/2},
\end{equation}
where the  pulse amplitude $\mathcal{A}_n (t) = \mathcal{A}$ for $t_n\leqslant t \leqslant t_n + \lambda$ and 0 otherwise, lasting for a duration $\lambda \ll \Delta t$, with $t_n = n \Delta t$ being the time at which the $n$th pulse is applied. Here, we only consider the $\pi$ pulses for our investigation, which means that the amplitude $\mathcal{A}$ and the duration $\lambda$ of a pulse satisfy $2\mathcal{A}\lambda = \pm \pi$.   It is not difficult to obtain the time evolution operator in the present of dynamical decoupling
pulses at time $t=2N\Delta t + \epsilon$
\begin{eqnarray}
\mathbb{U}(t)=\left\{ {\begin{array}{*{20}{c}}
	{{\mathcal{U}_{\rm{o}}}(\epsilon){{[{\mathcal{U}_{\rm{c}}}]}^N}\begin{array}{*{20}{c}}
		{}&{}&{}&{\begin{array}{*{20}{c}}
			{}
			\end{array}}
		\end{array}\begin{array}{*{20}{c}}
		{}&{}
		\end{array}0 \leqslant \epsilon < \Delta t} \\
	{{\mathcal{U}_{\rm{o}}}(\epsilon - \Delta t){\mathcal{U}_{\rm{p}}}(\lambda ){\mathcal{U}_{\rm{o}}}(\Delta t){{[{\mathcal{U}_{\rm{c}}}]}^N}\begin{array}{*{20}{c}}
		{}
		\end{array}\Delta t \leqslant \epsilon  < 2\Delta t}
	\end{array}} \right.
\end{eqnarray}
where $N=\left[t/(2\Delta t)\right]$ with $\left[ \dots \right]$  denoting the integer part, and the $\epsilon$ is the residual time after $N$ cycles. $\mathcal{U}_{\rm{o}}$ and  $\mathcal{U}_{\rm{p}}$ are the evolution operators corresponding to
the original Hamiltonian without and with the dynamical decoupling pulses, respectively. $\mathcal{U}_{\rm{c}}$ represents the  time evolution operator for an elementary cycle $2\Delta t$ which is given by $
\mathcal{U}_{\rm{c}} = \mathcal{U}_{\rm{p}} (\lambda)  \mathcal{U}_{\rm{o}} (\Delta t)   \mathcal{U}_{\rm{p}} (\lambda)  \mathcal{U}_{\rm{o}} (\Delta t)$.
For simplicity, we only focus on the periodic points $t_{2N}=2N\Delta t$. It has been found that  the decoherence factor $\Gamma(t)$ needs to be replaced by a new function $\Gamma_{\rm{p}} (N, \Delta t)$  in the presence of the decoupling pulses \cite{PhysRevA.69.030302},
\begin{eqnarray}
\Gamma_{\rm{p}} (N, \Delta t) =4 \int \rm{d}\omega J(\omega) \coth \left(\frac{\omega}{2T}\right)
 \times \frac{1-\cos (\omega t_{2N})}{\omega^2} \tan^2 \left(\frac{\omega \Delta t}{2}\right).
\end{eqnarray}
Comparing $\Gamma_{\rm{p}} (N, \Delta t)$ with $\Gamma (t)$, we notice that  the original bath spectral density has been transformed from $J(\omega)$ to the effective spectral density $J(\omega) \tan^2 (\omega \Delta t/2)$ after applying the bang-bang pulses.

In Fig. \ref{figure5}, we  illustrate the QSLT of qubit  versus the initial time  parameter $t$ with different  pulse intervals for the Ohmic spectrum  in the  strong-coupling regime. In the beginning of the evolution, the QSLT of qubit becomes shorter since the smaller pulse interval. On the contrary,  the  QSLT  decreases with the increase of pulse interval in the latter stage. Here, the fast  pulse is not only able to accelerate the dynamical evolution, but also  able to decelerate the dynamical evolution, which also plays a role of dual character. More interestingly, when $\Delta t$ is small enough (or in the limit $\Delta t\rightarrow 0$),  the bang-bang pulse enables us to
obtain a relative steady quantum evolution speed which almost remains constant. This is due to the fact that bang-bang pulse can effectively suppress the decoherence by averaging out the unwanted effects of environmental interaction\cite{PhysRevAbangbang,PhysRevA.69.030302,PhysRevLett.82.2417}.

\begin{figure}[tbph]
	\centering
	\subfigure[]{ \label{figure6a}
		\includegraphics[width=7cm]{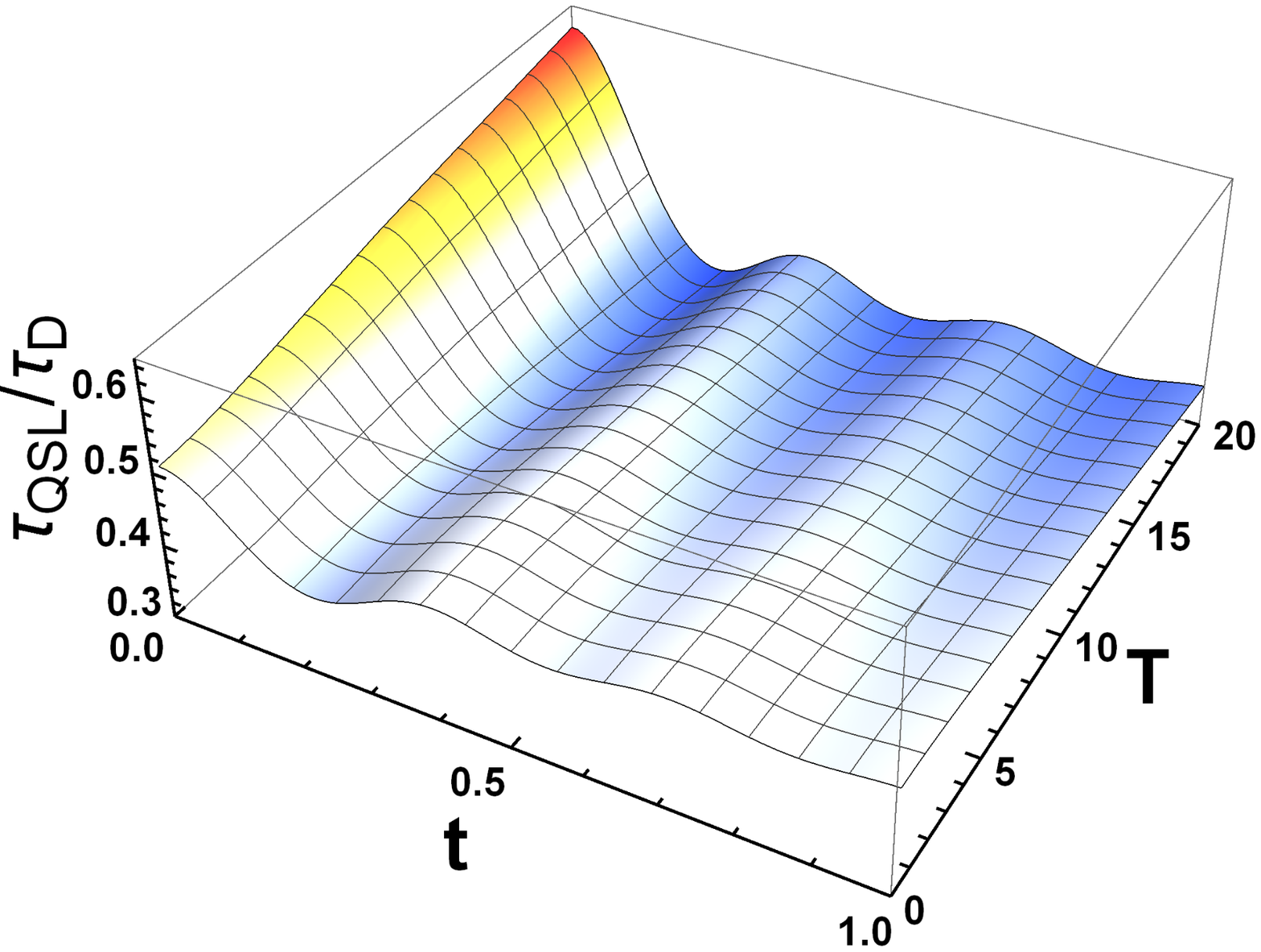}}
	\subfigure[]{ \label{figure6b}
		\includegraphics[width=7cm]{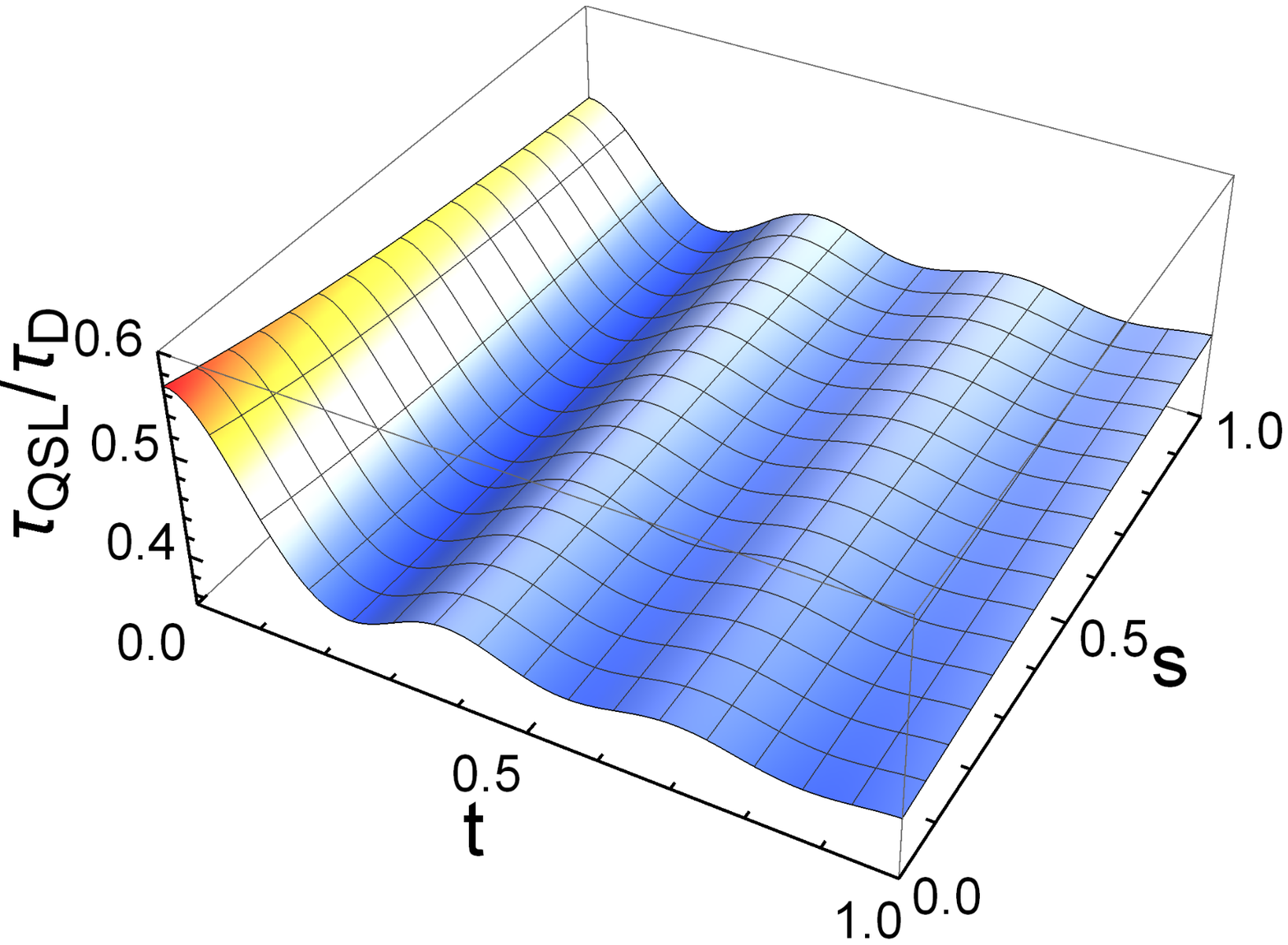}}
	\caption{(a)The QSLT $\tau_{\rm{QSL}}/\tau_{\rm{D}}$ of qubit   as a
		function of the initial time  parameter $t$ and bath temperature $T$ with bang-bang pulse for the sub-Ohmic spectrum $s=0.6$  in the strong-coupling regime  $\Lambda=0.2$. (b) The QSLT $\tau_{\rm{QSL}}/\tau_{\rm{D}}$ of qubit  as a function of the initial time  parameter $t$  and Ohmicity parameter $s$ with bang-bang pulse at bath temperature $T=1$ in the strong-coupling regime $\Lambda=0.2$. Other Parameters are chosen as $\tau_{\rm{D}}=1$, $\omega_{\rm{c}}=20$, $\Delta t =0.05$ and $\Omega =1$.  }
\end{figure}
Next, we turn to focus on the quantum evolution speed for a sub-Ohmic spectrum in the present of bang-bang pulse. We mainly investigate the effects of bath temperature $T$ [see Fig. \ref{figure6a}] and Ohmicity parameter $s$ [see Fig. \ref{figure6b}] on the QSLT  in the strong-coupling regime. Comparing to  Fig. \ref{figure4a}, we can observe from  Fig. \ref{figure6a} that   the bath temperature  plays a role of dual character in influencing the quantum  evolution speed in the strong-coupling regime since  the applied bang-bang pulse, which can not be found in the case without bang-bang pulse.  Figure \ref{figure6b} presents the QSLT as a function of the initial time  parameter $t$ and Ohmicity parameter $s$ with bang-bang pulse   in the strong-coupling regime.  In this situation, the Ohmicity parameter $s$ also plays a role of dual character in influencing the evolution of qubit, which is different in the case without bang-bang pulse [Fig. \ref{figure4c}]. In the beginning of evolution, the larger $s$ leads to a shorter QSLT which corresponds to a speed-up evolution.  After a certain time, instead, the larger $s$  leads to a longer QSLT which corresponds to a speed-down evolution.

In general, on the one hand, the bang-bang pulse can be used to control the quantum evolution speed in this dephasing model. On the other hand, since the presence of bang-bang pulse, the relevant
environmental parameters, such as bath temperature  and Ohmicity parameter, play  some more complicated and diverse roles in affecting the quantum evolution speed.

%%%%%%%%%%%%%%%%%%%%%%%%%%%%%%%%%%%%%%%%%%%%%%%%%%%%%%%%%%%%
\section{Quantum evolution speed in the nonlinear environment}\label{sec3}
Next, we consider another model: a bare qubit (labeled $A$) interacts with the other one (labeled $B$) which is coupled to a thermal bath. The qubit $B$ and thermal bath constitute the well-known spin-boson model which is used as a nonlinear environment of qubit $A$  \cite{PRL.120401}. The Hamiltonian of total system is given by
\begin{eqnarray}
\mathcal{H}^{\prime} &=&H_{\rm{s}}+H_{\rm{b}}+H_{\rm{int}}, \\
H_{\rm{s}}&=&\frac{\Omega_{\rm{A}}}{2} \sigma_z^{\rm{A}} +\frac{\Omega_{\rm{B}}}{2} \sigma_z^{\rm{B}}, \hspace{1cm} H_{\rm{b}}= \sum_{k} \omega_k b_k^\dagger b_k, \\
H_{\rm{int}}&=&f(s)g(b)+g_0\sigma_z^{\rm{A}}\sigma_z^{\rm{B}}
\end{eqnarray}
where   $H_{\rm{s}}$ and $H_{\rm{b}}$ are the Hamiltonians of two qubits and the bath, respectively. $g_0$ represents the interaction strength between the two qubits.  $f (s)=\sigma_z^{\rm{B}}$ is the subsystem’s operator coupled to its surrounding bath. The $g(b)= \sum_{k} g_k  ( b_k^\dagger + b_k)$  denotes the bath operator.
 We only focus on the  on-resonance case: $\Omega_{\rm{A}} = \Omega_{\rm{B}} = \Omega$. This model has been studied in the precious articles \cite{PRL.120401,Huang2010Effect},  however,  there are no exact analytical expression of the reduced density matrix for qubits, and  their results involve the Born-Markov approximation. Fortunately, by resorting to the HEOM method which is beyond the Born-Markov approximation,  we can deal with this model numerically.
  As a nonperturbative numerical method,  the  HEOM consists of a set of differential equations for the reduced subsystem and enables some rigorous studies in  chemical and biophysical systems,
  such as the optical line shapes of molecular aggregates \cite{1.3213013} and the quantum entanglement in photosynthetic light-harvesting complexes\cite{Sarovar2009Quantum}.
For the  finite-temperature case, we  consider the Ohmic spectrum with Drude cutoff:
\begin{equation}
	J(\omega) = \frac{2\Lambda \omega_{\rm{c}} \omega}{\pi(\omega_{\rm{c}}^2+\omega^2)},
\end{equation}
in which $\omega_{\rm{c}}$ is the cutoff frequency and $\Lambda$ represents the couping strength between the qubit and the bath. Then the bath correlation function $C(t)$ can be expressed as \cite{JPSJ.58.101,JPSJ.75.082001,PhysRevE.75.031107,1.2713104}
\begin{equation}
	C(t)=\sum_{k=0}^{\infty} \zeta_k e^{-\nu_k t},
\end{equation}
where the real and imaginary parts of $\zeta_k$ are respectively given as
\begin{eqnarray}
\zeta_k^{\rm{R}}&=& 4\Lambda \omega_{\rm{c}} T \frac{\nu_k}{\nu_k^2-\omega_{\rm{c}}^2}(1-\delta_{k0})+ \Lambda \omega_{\rm{c}} \cot(\frac{\omega_{\rm{c}}}{2T})\delta_{k0},\\
\zeta_k^{\rm{I}}&=& -\Lambda \omega_{\rm{c}} \delta_{k0},
\end{eqnarray}
with the $\nu_k=2k\pi T(1-\delta_{k0})+\omega_{\rm{c}} \delta_{k0}$ being the $k$-th  Matsubara frequency. Since the bath correlation function can be approximately expressed as the sum of the first few terms
in the series, the Matsubara frequency has been cut off and the convergence of result has been checked in our numerical calculation.

Following the derivation
shown in Ref. \cite{PhysRevA.85.062323}, the hierarchy equations of reduced quantum subsystem  can be  obtained as follows:
\begin{equation}
	\frac{\rm{d}}{\rm{d}t} \rho_{\vec{l}} (t) = -(i H_{\rm{s}}^{\times} + \vec{l} \cdot \vec{\nu})  \rho_{\vec{l}} (t) + \phi \sum_{k=0}^{\epsilon}  \rho_{\vec{l}+\vec{e}_k} (t) + \sum_{k=0}^{\epsilon} l_k \psi_k \rho_{\vec{l}-\vec{e}_k} (t),
\end{equation}
where $\vec{l}=(l_0, l_1,\ldots, l_{\epsilon})$ is a $(\epsilon+1)$-dimensional index, $\vec{\nu}= (\nu_0,\nu_1,\ldots,\nu_{\epsilon})$ and $\vec{e}_k =(0,0,\ldots,1_k,\ldots,0)$ are $(\epsilon+1)$-dimensional vectors with $\epsilon$ being the cutoff number of the Matsubara frequency. Two superoperators $\phi$ and $\psi_k$ are defined as
\begin{equation}
	\phi=i f(s)^\times,  \hspace{1cm} \psi_k=i \left[ \zeta_k^{\rm{R}}f(s)^\times +i \zeta_k^{\rm{I}} f(s)^\circ\right],
\end{equation}
with $X^\times Y=\left[X,Y\right]=XY-YX$ and $X^\circ Y= \left\lbrace X,Y\right\rbrace  =XY + YX$.

\begin{figure}[htbp]
	\centering
	\includegraphics[width=9cm]{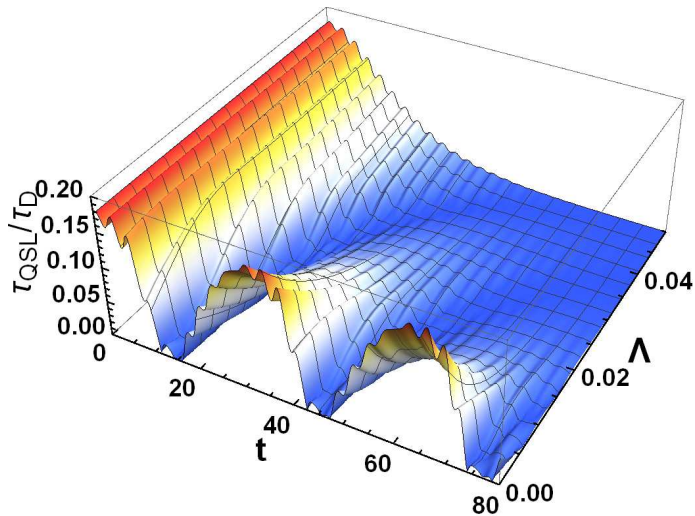}
	\caption{ The QSLT $\tau_{\rm{QSL}}/\tau_{\rm{D}}$ of qubit $A$  versus the initial time  parameter $t$ and  the coupling strength $\Lambda$ for the Ohmic spectrum. Other Parameters are chosen as $T=5$, $\tau_{\rm{D}}=10$, $g_0=0.1$,$\omega_{\rm{c}}=5$ and $\Omega =1$.}
	\label{figure7}
\end{figure}

We choose the initial state of two qubits as
\begin{equation}
\rho_{\rm{AB}}(0)=\frac{1}{2}{\left( {\begin{array}{*{20}{c}}
		1&1 \\
		1&1
		\end{array}} \right)_{\rm{A}}} \otimes \frac{1}{2}{\left( {\begin{array}{*{20}{c}}
		1&1 \\
		1&1
		\end{array}} \right)_{\rm{B}}},
\end{equation}
 and display the QSLT of qubit $A$ as function of the initial time parameter and the coupling strength for Ohmic spectrum in Fig. \ref{figure7}.
 In the strong-coupling regime, the quantum evolution  speed of  qubit  $A$  exhibits a speed-up process since the quantum decoherence effect.  In contrast, the QSLT of  qubit $A$ in the weak-coupling regime decreases to a minimum in the beginning of the evolution, then revivals and occurs a damped oscillatory  behavior.  As the coupling strength $\Lambda$ increases, the damped oscillatory  behavior fades away. The QSLT  behaves different in the strong-coupling and weak-coupling regimes.

\begin{figure}[t]
	\centering
	\subfigure[]{\label{figure8a}
		\includegraphics[width=7cm]{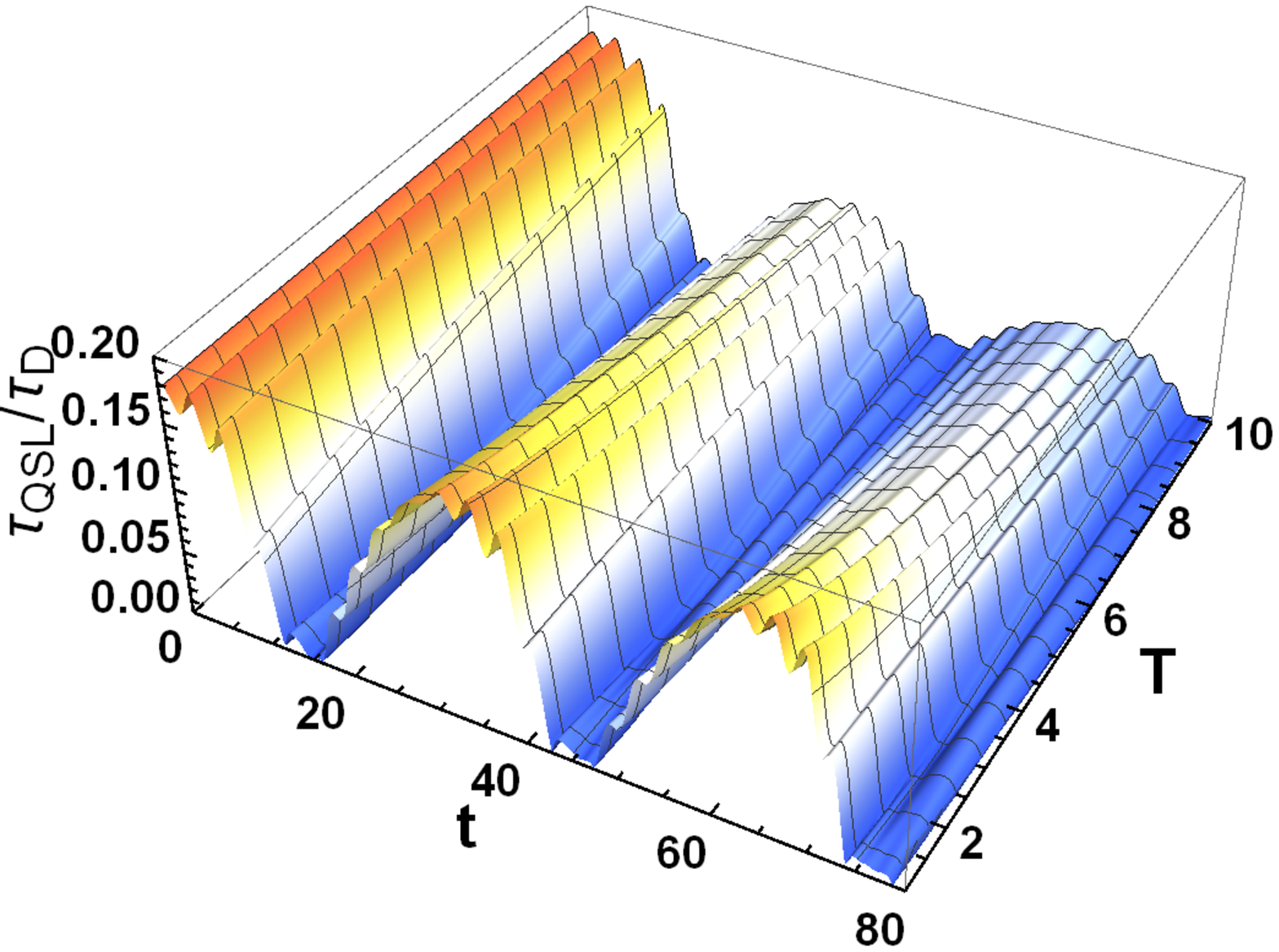}}
	\subfigure[]{\label{figure8b}
		\includegraphics[width=7cm]{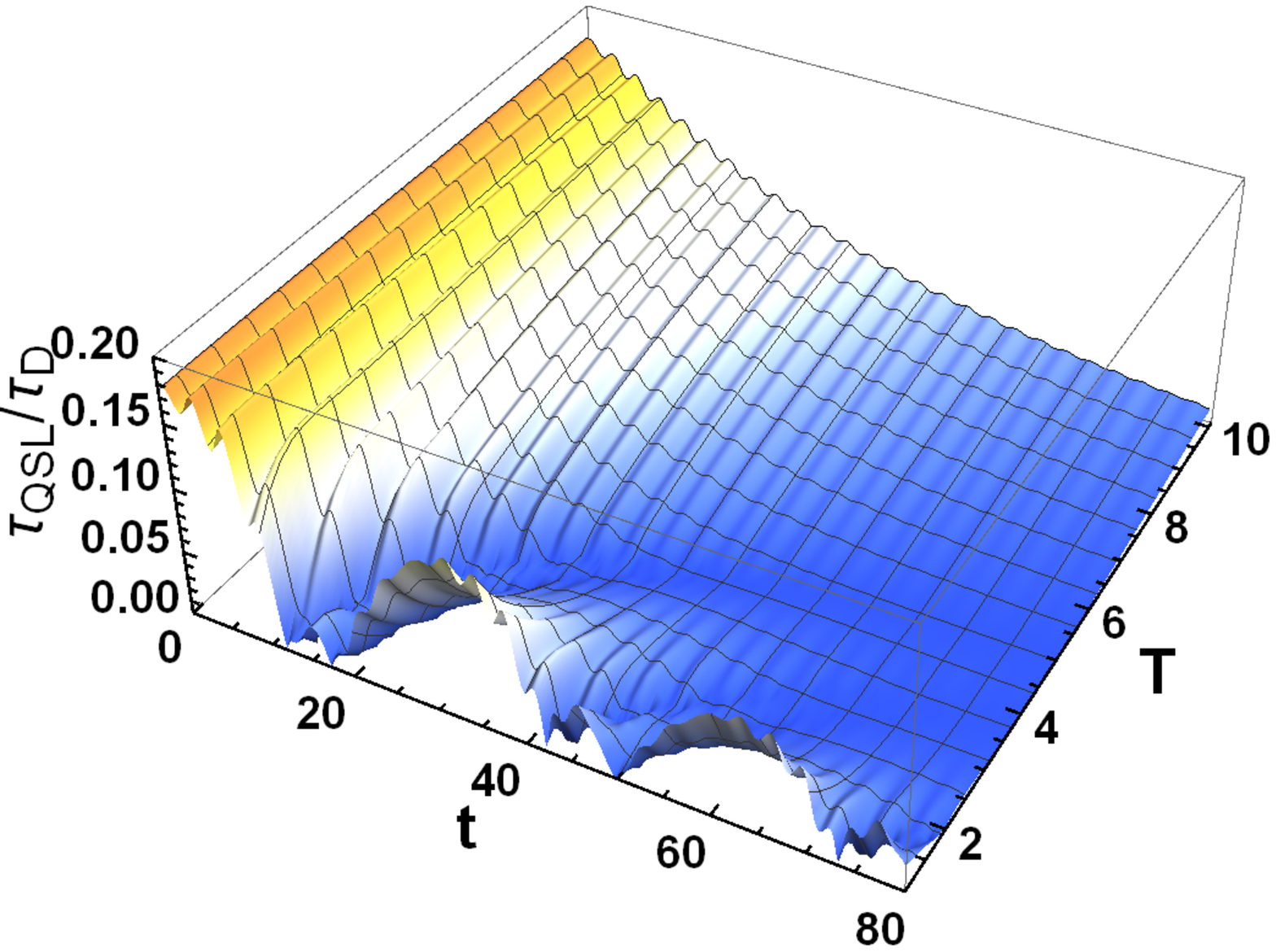}}
\subfigure[]{\label{figure8c}
		\includegraphics[width=7cm]{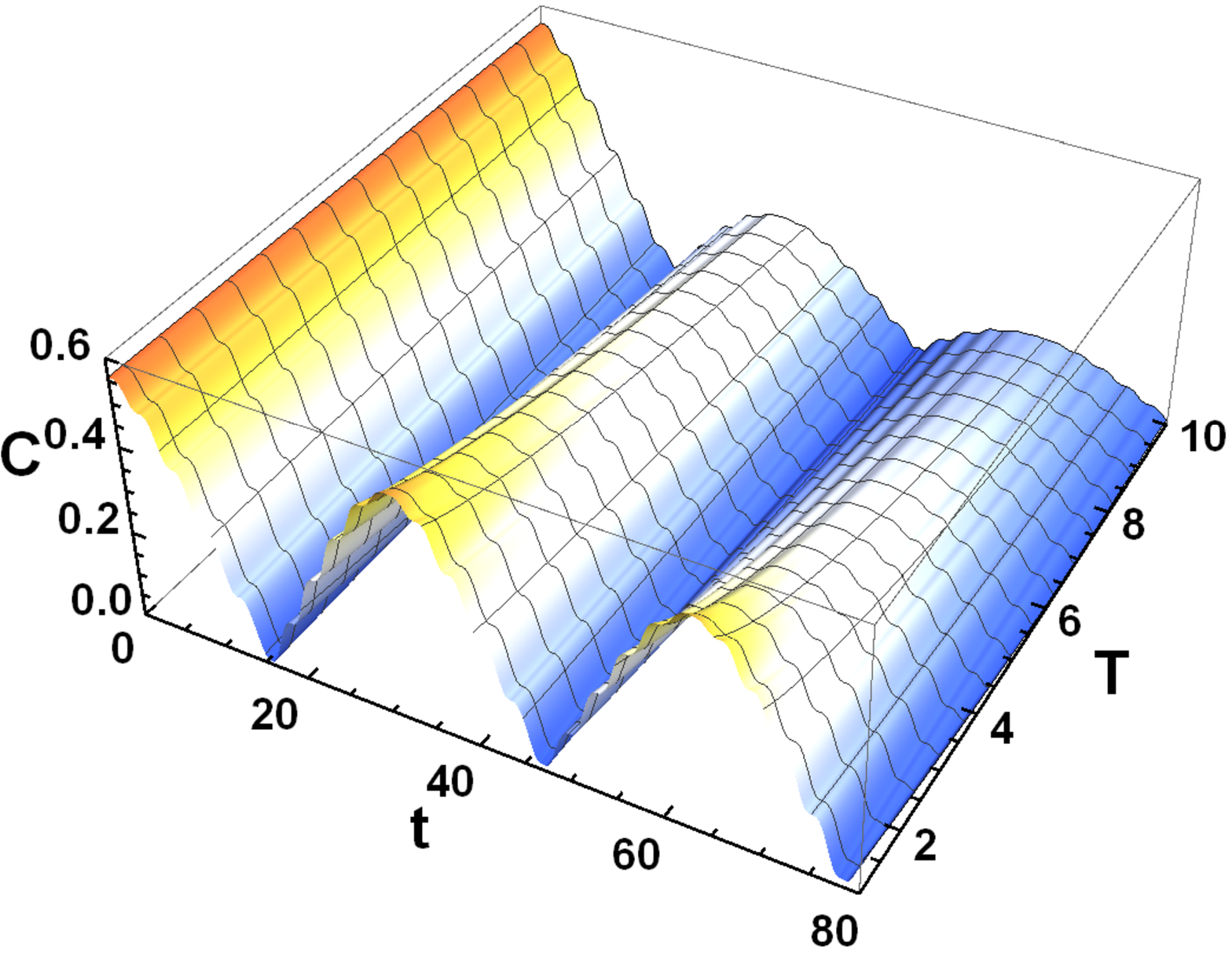}}
\subfigure[]{\label{figure8d}
		\includegraphics[width=7cm]{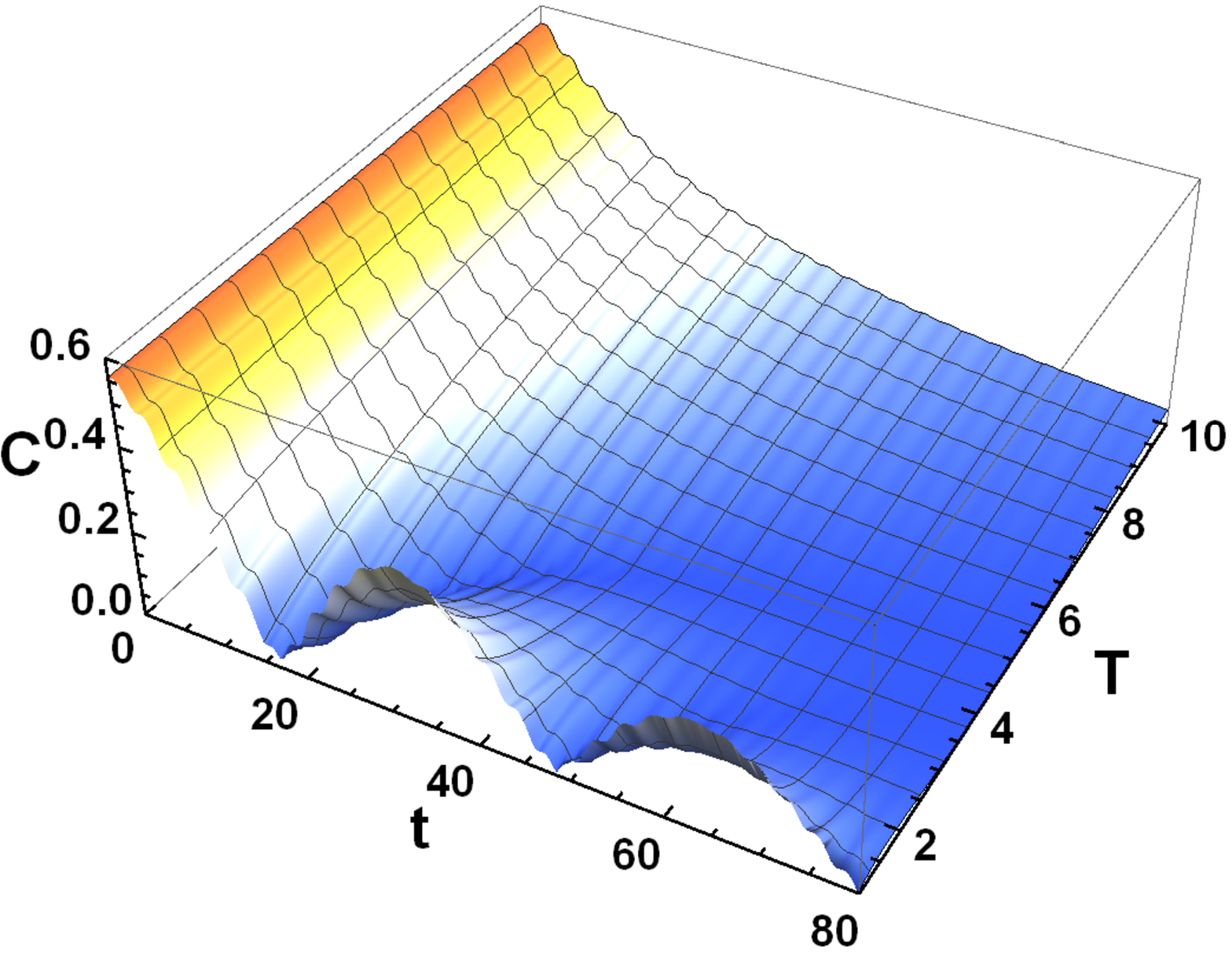}}
	\caption{The QSLT $\tau_{\rm{QSL}}/\tau_{\rm{D}}$ and quantum coherence $C$ of qubit $A$  versus the initial time  parameter $t$ and  the temperature $T$ for the (a)(c) weak-coupling $\Lambda=0.005$ regime and (b)(d) strong-coupling $\Lambda=0.05$ regime. Other Parameters are chosen as $\tau_{\rm{D}}=10$, $g_0=0.1$,$\omega_{\rm{c}}=5$ and $\Omega =1$.}
\label{figure8}
\end{figure}

In the previous section, it is shown that the quantum evolution speed is related to the dynamics of quantum coherence. Here, we choose the measure of quantum coherence  based on the quantum Jensen-Shannon divergence  \cite{Radhakrishnan2016} to study the dynamics of quantum coherence of  qubit $A$ for getting more insight on the QSLT. The expression of quantum coherence is given by
\begin{equation}
  C(\rho)=\sqrt{S\left(\frac{\rho+\rho_{\rm{dia}}}{2}\right)-\frac{S(\rho)+S(\rho_{\rm{dia}})}{2}}
  \end{equation}
  where $S(\rho)=-\rm{Tr}\rho \log_2 \rho $ is the von Neumann entropy and $\rho_{\rm{dia}}$ is the incoherent state obtained from $\rho$ by deleting all off-diagonal elements\cite{PhysRevLett.Coherence}.

We plot the QSLT and quantum coherence of qubit $A$ as functions of the  initial time  parameter $t$ and  the temperature $T$  in Figs.\ref{figure8}. In the weak-coupling regime, as shown in Figs.\ref{figure8a} and \ref{figure8c}, the QSLT and quantum coherence exhibit the  damped oscillatory  behaviors and  have similar evolutions.
The increase of temperature induces the speed-up evolution  since the fact that higher temperature brings more intensive decoherence, which can also be confirmed by the dynamics of quantum coherence in Fig. \ref{figure8c}.  In contrast, there are some rich and anomalous phenomenons in the strong-coupling regime as shown in Figs.\ref{figure8b} and \ref{figure8d}. In the low-temperature region, the behaviors of QSLT and quantum coherence are analogous to those  in the weak-coupling regime. However, as the temperature increases, the QSLT is extended which means the quantum evolution speed is decelerated and not a  monotonic increasing function of the bath temperature any more. Superficially, this result is due to the enhancement of quantum coherence by the high temperature,  which can be confirmed in the dynamics of quantum coherence in Fig. \ref{figure8d}. The underlying reason for this  anomalous phenomenon is that the spin-boson model consisting of qubit $B$ and the thermal bath can be seen as a nonlinear environment for qubit $A$, the power spectrum doesn't necessarily grow   with the  temperature \cite{PRL.120401}.  Thus, some reversed effects occur, i.e., the increase of bath temperature may give rise to the speed-down evolution and the enhancement of quantum coherence.

%%%%%%%%%%%%%%%%%%%%%%%%%%%%%%%%%%%%%%%%%%%%%%%%%%%%
\section{Conclusion}\label{sec4}

In conclusion, we  have considered two kinds of  finite-temperature bosonic  baths to investigate the  quantum evolution speed of qubit and find that the  quantum evolution speed isn't a monotonic function of temperature.  For the spin-boson model, the quantum evolution speed of qubit  can be accelerated by the high temperature in the strong-coupling regime. In the weak-coupling regime, the bath temperature plays a role of dual character in affecting the  quantum evolution speed, which means that the high  temperature not only leads to the speed-up  but also  speed-down processes.  The quantum coherence, relative purity and the driving time are responsible for the different behaviors of quantum evolution speed  in the strong-coupling and weak-coupling regimes. Furthermore, we can observe that the quantum evolution speed can be controlled by the bang-bang pulse  in the strong-coupling regime, and  the relative steady  quantum evolution speed can be obtained by fast pulse. Interestingly, the bath temperature and  Ohmicity parameter also play roles of dual character in the strong-coupling regime since the presence of bang-bang pulse, which are not found in the case without pulse.  For the nonlinear bath, we study the quantum  evolution speed of qubit by applying the hierarchical equations of motion method. It is  shown that
the performances of quantum  evolution speed in  weak-coupling and strong-coupling regimes are very different. In the strong-coupling situation, the  quantum evolution speed  at low-temperature region
behaves similarly to that in the weak-coupling situation where the  quantum evolution speed is only a monotonic increasing function of temperature. However, the rise of temperature induces the speed-down process, this anomalous phenomenon is  on account of the   temperature dependence of the spectral profile in nonlinear bath.  As a comparison, the dynamics of quantum coherence is also explored in different situations.   These results provide the possibilities to control quantum evolution speed by changing the relevant  environmental parameters in the finite-temperature bosonic environments. Finally, we expect our studies to be of interest for experimental applications in quantum computation and quantum information processing.
%%%%%%%%%%%%%%%%%%%%%%%%%%%%%%%%%%%%%%%%%%%%%%%%%%%%%%%%%%%%%%%%%
\section{Acknowledgments}

This project was supported by the National Natural Science Foundation of China (Grant No.11274274) and the Fundamental Research Funds for the Central Universities (Grant No.2017FZA3005 and 2016XZZX002-01).

\section*{References}

\bibliographystyle{elsarticle-num}
%\bibliography{tddmrg}
%%%%%%%%%%%%%%%%%%%%%%%%%%%%%%%%%%%%%%%%%%%%%%%%%%%%%%%%%%%%%%%%%%%%%%%%%%%%

\end{document}